\documentclass[aip,pof,amsmath,amssymb,reprint,onecolumn,longbibliography]{revtex4-1}

\usepackage{graphicx}
\usepackage[percent]{overpic}
\usepackage{bm}

\usepackage[utf8]{inputenc}
\usepackage[T1]{fontenc}
\usepackage{mathptmx}
\usepackage{etoolbox}
\usepackage{booktabs}

\usepackage{xcolor}

\makeatletter
\def\@email#1#2{%
 \endgroup
 \patchcmd{\titleblock@produce}
  {\frontmatter@RRAPformat}
  {\frontmatter@RRAPformat{\produce@RRAP{*#1\href{mailto:#2}{#2}}}\frontmatter@RRAPformat}
  {}{}
}%
\makeatother
\begin{document}

\title[Identifiability Limits of Reynolds-Stress Attribution]
{Identifiability Limits of Reynolds-Stress Attribution in
Localized Nonorthogonal Flow Representations}

\author{Peng Cheng}
 \email{peng.cheng.hit@gmail.com}
\author{Ming Jian}%
\author{Wen Tong}
\affiliation{Huawei Technologies Canada Co., Ltd., \\ 303 Terry Fox Drive, Suite 101, Ottawa, ON, K2K 3J1, Canada}

\date{\today}

\begin{abstract}
Reynolds stress is assembled from correlations among velocity fluctuations;
assigning it to named vortical structures additionally requires their velocity
fields to be separable.
We examine this requirement in homogeneous isotropic turbulence from the
Johns Hopkins Turbulence Databases (JHTDB) using a frozen solenoidal expansion of curl-lifted
anisotropic Gaussian vorticity functions.
The expansion retains locally addressable blob-, sheet-, and tube-like geometry,
while its fixed Biot--Savart map carries coefficient covariance exactly into the stress of the
represented Gaussian velocity.
Principal-angle geometry and self--cross
covariance then connect overlap between group velocity spans to stress
assembly.
At a principal cosine of \(0.95\), \(55\%\) of the sheet-associated
directions are also closely represented by the tube-like family, and products
within the shared directions carry \(99.6\%\) of the covariance cancellation.
Removing between-family covariance increases the direct numerical simulation (DNS)-referenced stress error
by factors of \(53.6\)--\(87\).
Orthogonal Fourier and proper orthogonal decomposition (POD) regroupings of the same velocity remove this cancellation,
showing that it follows from the local nonorthogonal organization rather than from the represented field alone.
The morphology labels remain representation-dependent, but their physical
consequence is direct: Reynolds stress in this local description is carried by
coupled velocity fluctuations, not by additive component energies.
Localized nonorthogonal representations can therefore reveal how turbulent momentum
flux is assembled, provided that span overlap and cross covariance are retained.
\end{abstract}
\maketitle

\section{Introduction}
\label{sec:introduction}

Reynolds stress is a covariance of velocity fluctuations. It contains variance
within a class of motions and covariance between classes. Once a turbulent
field is decomposed into named vortical components, attributing stress to one
class therefore requires more than accurate reconstruction: the component
velocity fields must also be separable. Reconstruction constrains their sum,
whereas second-order attribution depends on how that sum is partitioned. In an
overcomplete nonorthogonal expansion, different coefficient vectors can
generate the same velocity field \cite{christensen2016frames}; uniqueness is
then controlled by the geometry of the expansion, not by approximation error
alone \cite{donoho2003sparse,gribonval2003unions,tropp2004greed,
elad2010sparse}. Related ambiguity arises when modal coefficients do not retain
separable meaning \cite{callaham2022nonlinear,mendez2019mpod,
murata2020nonlinearmodes}. Component-wise Reynolds-stress attribution thus
requires its own geometric and statistical test.

Orthogonal and localized representations answer different physical questions.
Proper orthogonal decomposition (POD) is energy-optimal at fixed rank
\cite{lumley1967pod,sirovich1987snapshot,berkooz1993proper,
holmes2012turbulence}, while Leray-projected Fourier modes provide an accurate
global coordinate for periodic homogeneous flow
\cite{canuto2006spectral,orszag1972homogeneous,rogallo1981homogeneous}.
Localized frames instead make intermittent rotational content spatially
addressable, as in wavelet descriptions of turbulence
\cite{farge1992wavelets,farge2001cvs,schneider2010wavelet} and solenoidal
wavelet constructions \cite{deriaz2005divfreewavelets,
lessig2018divfreepolar}. Locality and morphological flexibility can therefore
be scientifically useful even when they introduce overlap. The relevant
question is which component statements survive that overlap, not whether a
local representation should simply be replaced by a global orthogonal basis.

We study this tradeoff with a fixed, data-adapted expansion of vorticity in
directional Gaussian functions. Their centers, widths, and orientations give
locally addressable blob-, sheet-, and tube-like support. Curl lifting makes
each sampled vorticity function solenoidal, and a fixed Biot--Savart operator
maps its amplitude to incompressible velocity \cite{majda2002vorticity}. A
velocity--vorticity graph norm couples local rotational fidelity to the decoded
velocity, while Gram regularization limits nearly duplicate directions during
geometry determination. The geometry and fitting convention are frozen before
new snapshots are analyzed, so each coefficient retains the same velocity
meaning across an ensemble. Related Gaussian constructions appear in
radial-basis approximation and data-adapted flow representations
\cite{buhmann2003radial,moody1989rbf,ni2025divfree,
shenoy2026gaussianfield,xing2025gaussian,xing2026ddfk}.

The frozen velocity map makes the second-order accounting exact. If centered
amplitudes have covariance \(C_A\) and the decoded velocity functions are
collected in \(\Phi\), then
\(R^{\mathrm{coef}}=\Phi C_A\Phi^{\mathsf T}\) is the Reynolds stress of
the represented Gaussian velocity. Agreement with direct numerical simulation (DNS) stress remains a
separate fidelity question. Partitioning \(\Phi\) into morphological families
splits the same tensor into self and cross covariance blocks. All blocks are
required for exact recovery, but overlapping group spans can make their
individual allocation nonunique. Stress attribution in a local nonorthogonal
representation is therefore a problem of coupled covariance, not a
nonnegative census of isolated structures.

We connect that covariance balance to representation geometry through four
measurements. Gram conditioning quantifies dependence across the complete
expansion; principal angles locate directions represented nearly
interchangeably by different groups; signed self--cross alignment measures the
cancellation hidden in the assembled stress; and paired reallocations probe
component changes at fixed field or fixed group membership. These tests define
identifiability only for the declared representation, grouping, inner product,
finite ensemble, and perturbation family. Passing such a test is necessary,
but not sufficient, for assigning physical meaning to a component; it does not
establish causality or uniqueness under another representation.

The homogeneous isotropic turbulence (HIT) results reveal a direct connection
between span overlap and stress assembly. At principal cosine \(0.95\),
\(55\%\) of the sheet-associated velocity directions are also closely
represented by the tube-like family, and products within those shared
directions carry \(99.6\%\) of the covariance cancellation. Removing cross
covariance raises the DNS-referenced stress error by more than a factor of
fifty. Orthogonal regrouping of the same velocity and covariance controls show
that the effect belongs to the nonorthogonal dictionary--covariance pair, not
to the represented field alone. The channel-flow extension gives the
complementary inhomogeneous result: retaining the magnitude of Reynolds shear
does not recover production when the shear is displaced from the mean-gradient
region. Together, these results separate stress fidelity, spatial placement,
and component separability while showing why all three matter for a physical
description of turbulent momentum transport.

Section~\ref{sec:method} defines the solenoidal representation, covariance
decomposition, and separability diagnostics. Section~\ref{sec:coordinate-verification}
selects the frozen geometry. Sections~\ref{sec:jhtdb-attribution} and
\ref{sec:channel-extension} examine Reynolds-stress structure in homogeneous
isotropic turbulence and channel flow. Sections~\ref{sec:discussion} and
\ref{sec:conclusion} summarize the physical implications and scope.

\section{Solenoidal representation and covariance structure}
\label{sec:method}

The method fixes a locally anisotropic, solenoidal expansion before analyzing
Reynolds stress. Gaussian functions describe localized vorticity geometry,
curl lifting enforces solenoidality, and the Biot--Savart map gives
incompressible velocity. Once the geometry and amplitude-fitting convention
are fixed, coefficient covariance determines the Reynolds stress of the
Gaussian velocity contribution exactly. Morphological group spans and their
self--cross covariance blocks then show whether that stress can be separated
among groups. Section~S1 of the supplementary material lists the notation and
its defining equations.

\subsection{Solenoidal vorticity and velocity recovery}
\label{subsec:vorticity-decoding}

On \(\Omega=\mathbb T^3\), let \(U\in\mathbb R^{S\times N_u}\) collect the
vectorized velocity snapshots \(\mathbf u_s\), \(s=1,\ldots,S\), and let
\(Y\in\mathbb R^{S\times N_\omega}\) collect their vorticities
\(\boldsymbol\omega_s=\nabla_h\times\mathbf u_s\). Vorticity exposes
localized, derivative-bearing rotational content while retaining a linear
route to incompressible velocity. In a storage-matched two-dimensional
comparison, it gives lower velocity, vorticity, and velocity-gradient errors
than direct velocity and streamfunction representations at every tested budget
(supplementary material, Sec.~S2).

For envelope geometry \(\theta\), let \(G_\theta\) denote the Gaussian
vorticity synthesis operator defined in
Sec.~\ref{subsec:gaussian-vorticity-atoms}; its discretization
\(G_\theta\in\mathbb R^{N_\omega\times M_b}\) is the dictionary used below.
Once geometry determination fixes
\(\theta^\star\) (Sec.~\ref{subsec:graph-norm-fitting}), we write
\(G=G_{\theta^\star}\), apply \(\mathcal K\) columnwise, and freeze the decoded
velocity dictionary \(\Phi=\mathcal K G\). If
\(A\in\mathbb R^{S\times M_b}\) collects the snapshot-dependent amplitudes,
the represented vorticity and velocity ensembles are
\begin{equation}
  \widehat Y=AG^{\mathsf T},
  \qquad
  \widehat U^{G}=A\Phi^{\mathsf T}.
  \label{eq:representation-decoder}
\end{equation}
The same amplitude matrix therefore represents vorticity and its
decoded Gaussian-channel velocity.

The fixed decoder \(\mathcal K\) acts mode by mode. Let \(\mathbf k\) denote
the Fourier wavevector and \(\mathbf k_d\) the Nyquist-zeroed derivative
wavevector used for both the snapshot curl and the Gaussian curl lift.
Retaining the full \(|\mathbf k|^2\) in the inverse-Laplacian Green operator
gives
\begin{equation}
  \widehat{\mathcal K\boldsymbol\omega}(\mathbf k)
  =
  \frac{
    i\mathbf k_d\times
    \widehat{\boldsymbol\omega}(\mathbf k)
  }{|\mathbf k|^2},
  \qquad \mathbf k\ne\mathbf0 .
  \label{eq:biot-savart-decoder}
\end{equation}
The zero mode is set to zero. Hence \(\mathcal K\)
returns a zero-mean, discretely divergence-free velocity and is an exact
inverse curl on the compatible subspace where \(\mathbf k_d=\mathbf k\).

In the periodic Hodge decomposition, vorticity determines the zero-mean
solenoidal velocity but not its spatially constant harmonic mode. Let
\(\overline U_h\) collect the snapshot-wise spatial means
\(\overline{\mathbf u}_{s,h}=\langle\mathbf u_s\rangle_h\), broadcast over the
grid. Restoring this exact channel gives the full represented ensemble
\begin{equation}
  \widehat U
  =
  \overline U_h+\widehat U^G .
  \label{eq:hodge-deployment}
\end{equation}
The three exact mean-velocity components enter fidelity evaluation but are
not geometric amplitudes and receive no Gaussian-envelope label.

The remaining construction concerns the localized vorticity functions
whose decoded velocities form the nonconstant channel.

\subsection{Directional Gaussian vorticity functions}
\label{subsec:gaussian-vorticity-atoms}

The atom construction separates adaptable local geometry from kinematic
admissibility. Scalar Gaussian envelopes determine where, and over what local
shape, rotational content is represented; the curl lift maps that geometry
into the solenoidal vorticity space required by the decoder. Neither
constraint is inferred from data or imposed by a penalty.

For prescribed center \(\boldsymbol\mu_i\), the sampled envelope is
\begin{equation}
  g_i^\theta(\mathbf x)
  =
  \exp\!\left[
    -\frac12
    \mathbf d_i(\mathbf x)^{\mathsf T}
    \Sigma_i^{-1}
    \mathbf d_i(\mathbf x)
  \right],
  \label{eq:gaussian-geometry}
\end{equation}
where \(\mathbf d_i\) wraps each Cartesian displacement into the nearest
periodic image. We consider fixed isotropic
\(\Sigma_i=\sigma_0^2\mathbf I\), learned isotropic
\(\Sigma_i=\sigma_i^2\mathbf I\), axis-aligned anisotropic
\(\Sigma_i=\operatorname{diag}(\sigma_{i,1}^2,\sigma_{i,2}^2,
\sigma_{i,3}^2)\), and freely oriented anisotropic
\(\Sigma_i=Q_i\operatorname{diag}(\sigma_{i,1}^2,\sigma_{i,2}^2,
\sigma_{i,3}^2)Q_i^{\mathsf T}\), with \(Q_i\in SO(3)\). Centers are fixed;
widths and, for the last family, rotations are inferred. Here isotropy describes
the scalar envelope rather than the turbulence statistics.

Anisotropic widths let the local support adapt to elongated, directionally
organized rotational content. Independent geometry determination resolves a
stress-fidelity gain from axis-aligned anisotropic widths over learned
isotropic widths, but no additional free-orientation gain under the retained
coherence control at the tested budget. We retain the oriented parameterization
in the frozen representation.
Appendix~\ref{app:geometry-family-evidence} reports this comparison and its
conditioning tradeoff.

Let \(\mathcal N_W\mathbf v=\mathbf v/\|\mathbf v\|_W\) denote normalization
in the sampled fitting metric defined below. Inserting envelope
\(g_i^\theta\) into Cartesian seed \(\mathbf e_c\) gives the vorticity atom
\(\mathbf g_{i,c}^\theta=
\mathcal N_W[\nabla_h\times(g_i^\theta\mathbf e_c)]\), \(c=1,2,3\).
The complete atom family defines the synthesis operator
\begin{equation}
  G_\theta:
  \mathbb R^{M_b}
  \longrightarrow
  \mathcal V_{\omega,h}^{0,\mathrm{div}},
  \qquad
  G_\theta\mathbf a
  =
  \sum_{i=1}^{N_G}\sum_{c=1}^{3}
  a_{i,c}\mathbf g_{i,c}^{\theta},
  \qquad
  M_b=3N_G .
  \label{eq:curl-lifted-atoms}
\end{equation}
Here \(\mathcal V_{\omega,h}^{0,\mathrm{div}}\) is the discrete zero-mean,
solenoidal vorticity space. Because the divergence and spatial mean of a
periodic curl vanish, the range of \(G_\theta\) lies in this admissible space
and is compatible with \(\mathcal K\). Its matrix discretization has one column
per envelope--channel pair. Envelope preprocessing and normalization details
are given in Appendix~\ref{app:reproducibility}.

With the admissible atom family fixed, geometry determination can be
separated from snapshot-dependent amplitude inference.

\subsection{Frozen geometry and amplitude inference}
\label{subsec:graph-norm-fitting}

Geometry learning uses variable projection: amplitudes are solved analytically
for each candidate geometry. For
fixed \(\theta\), the ridge amplitude matrix
\(A_\theta\in\mathbb R^{S\times M_b}\) is
\begin{equation}
  A_\theta
  =
  YWG_\theta
  \left(
    G_\theta^{\mathsf T}WG_\theta
    +\lambda_{\mathrm{ridge}}\mathbf I
  \right)^{-1},
  \qquad \lambda_{\mathrm{ridge}}\ge0,
  \label{eq:amplitude-fit}
\end{equation}
where \(W\succeq0\) is the sampled fitting metric. Its target full-grid
velocity--vorticity graph norm is
\begin{equation}
  \begin{aligned}
    \|\mathbf e_\omega\|_{\mathrm{gr},\ell}^{2}
    &=
    \sum_{\mathbf k\ne\mathbf0}
    w_\ell(\mathbf k)
    \|\widehat{\mathbf e}_\omega(\mathbf k)\|_2^2,\\
    w_\ell(\mathbf k)
    &=
    |\mathbf k|^{-2}+\ell^2,
    \qquad \ell>0 .
  \end{aligned}
  \label{eq:graph-norm}
\end{equation}
The first term equals the decoder-induced velocity metric on Nyquist-free
solenoidal modes; the second retains vorticity fidelity. In the implementation,
\(W=H_\ell^{\mathsf T}\mathcal S_h^{\mathsf T}\mathcal S_hH_\ell\), where
\(H_\ell\) has Fourier multiplier \(w_\ell^{1/2}\) and
\(\mathcal S_h\) is the frozen deterministic grid sampler. The
vorticity-\(L^2\) control sets \(H_\ell=\mathbf I\).

Substituting Eq.~\eqref{eq:amplitude-fit} eliminates the amplitudes from the
outer problem and determines the geometry from
\begin{equation}
  \theta^\star
  =
  \operatorname*{arg\,min}_{\theta}
  \frac{
    \|Y-A_\theta G_\theta^{\mathsf T}\|_W^2
    +\lambda_{\mathrm{ridge}}\|A_\theta\|_F^2
  }{\|Y\|_W^2}
  +\mathcal R_{\mathrm{geom}}(\theta)
  +\mathcal R_{\mathrm{Gram}}(\theta),
  \label{eq:learning-objective}
\end{equation}
where
\(\|Z\|_W^2=\operatorname{tr}(ZWZ^{\mathsf T})\).
Here \(\mathcal R_{\mathrm{geom}}=\iota_{\mathcal A}\) is the indicator of the
admissible width--aspect set \(\mathcal A\), enforced numerically by
projection. The
coherence term appearing in Eq.~\eqref{eq:learning-objective} is
\begin{equation}
  \mathcal R_{\mathrm{Gram}}(\theta)
  =
  \alpha\operatorname*{mean}_{m<n}
  \left[
    \max\!\left\{
      0,|(C_\theta)_{mn}|-\tau_{\mathrm{Gram}}
    \right\}
  \right]^2 ,
  \label{eq:gram-regularization}
\end{equation}
where
\(C_\theta=D_\theta^{-1/2}K_\theta D_\theta^{-1/2}\) is the normalized
graph-metric correlation matrix,
\(K_\theta=G_\theta^{\mathsf T}WG_\theta\) is the corresponding Gram matrix,
and \(D_\theta=\operatorname{diag}(K_\theta)\) contains its column
self-products.
With \(\alpha\ge0\) and \(0\le\tau_{\mathrm{Gram}}\le1\), the threshold permits local overlap
while penalizing excessive coherence and near-duplicate directions.

After optimization, \(G=G_{\theta^\star}\), the decoder \(\Phi\), and all envelope
labels are frozen; their normalized correlation and Gram matrices are denoted
\(C_G=C_{\theta^\star}\) and \(K_G=K_{\theta^\star}\), respectively. New
ensembles replace
\(Y\) in Eq.~\eqref{eq:amplitude-fit} with \(\theta=\theta^\star\); only
amplitudes are refitted. Numerical details are recorded in
Appendix~\ref{app:reproducibility}.

This frozen-map convention makes coefficient covariance comparable
across snapshots and permits an exact stress propagation.

\subsection{Reynolds stress from coefficient covariance}
\label{subsec:covariance-lift}

Let \(\Phi(\mathbf x)\in\mathbb R^{3\times M_b}\) collect the decoded velocity
atoms. Center the amplitudes in matrix form as
\(A'=P_SA\), where
\(P_S=\mathbf I_S-S^{-1}\mathbf 1_S\mathbf 1_S^{\mathsf T}\). The
Gaussian-channel Reynolds stress is
\begin{equation}
  R^{\mathrm{coef}}(\mathbf x)
  =
  \Phi(\mathbf x)C_A\Phi(\mathbf x)^{\mathsf T},
  \qquad
  C_A=\frac{1}{S-1}A'^{\mathsf T}A' .
  \label{eq:covariance-lift}
\end{equation}
By construction,
\(R^{\mathrm{coef}}=\operatorname{Cov}_s(\widehat{\mathbf u}^{G}_s)\)
exactly for the represented Gaussian velocity channel; its discrepancy from
the DNS sample stress is representation error.

Restoring the exact mean-velocity channel gives
\begin{equation}
  R^{\mathrm{dep}}(\mathbf x)
  =
  \operatorname{Cov}_s\!\left(
    \widehat{\mathbf u}_s(\mathbf x)
  \right).
  \label{eq:deployed-diagnostics}
\end{equation}
For \(X\in\{\mathrm{coef},\mathrm{dep}\}\), define
\(k^X=\tfrac12\operatorname{tr}R^X\) and
\(q^X=-\nabla_h\cdot R^X\).
Full-field quantities serve fidelity tests;
\(R^{\mathrm{coef}}\) serves geometric accounting.

The resulting tensor is exact for the represented Gaussian velocity;
the next question is whether a partition of that tensor separates its named
groups.

\subsection{Group covariance and span separability}
\label{subsec:group-attribution}

A fixed partition of the decoded velocity functions gives an exact covariance
decomposition. Whether that decomposition separates the named groups is a
different question, because distinct groups may span nearly the same velocity
directions. Let \(\{\mathcal G_p\}_{p=1}^{P_g}\) partition the frozen Gaussian
columns, with each envelope label inherited by its three curl channels. If
\(\Phi_p\) is the corresponding decoder block and \(C_{pq}\) the matching block
of \(C_A\), the ordered pair contribution is
\begin{equation}
  R^{(p,q)}(\mathbf x)
  =
  \Phi_p(\mathbf x)C_{pq}\Phi_q(\mathbf x)^{\mathsf T}.
  \label{eq:group-pair-accounting}
\end{equation}
We use the nonduplicated symmetric pair
\(\widetilde R^{(p,p)}=R^{(p,p)}\) and
\(\widetilde R^{(p,q)}=R^{(p,q)}+R^{(q,p)}\) for \(p<q\), so that
\(R^{\mathrm{coef}}=\sum_{p\le q}\widetilde R^{(p,q)}\).
Its signed alignment with the assembled stress and the resulting cancellation
index are
\begin{equation}
  \begin{aligned}
    \eta_{pq}^{R}
    &=
    \frac{
      \langle\widetilde R^{(p,q)},R^{\mathrm{coef}}\rangle_{L_F^2}
    }{
      \|R^{\mathrm{coef}}\|_{L_F^2}^{2}
    },
    \qquad
    \sum_{p\le q}\eta_{pq}^{R}=1,\\
    \mathcal C_R
    &=
    \sum_{p\le q}|\eta_{pq}^{R}|,
  \end{aligned}
  \label{eq:pair-alignment}
\end{equation}
where \(\langle\cdot,\cdot\rangle_{L_F^2}\) is the domain-integrated tensor
Frobenius inner product. Thus \(\eta_{pq}^{R}\) measures signed covariance
alignment, whereas \(\mathcal C_R\) measures the opposing contributions hidden
by their unit sum. Neither quantity is an ownership fraction.

Assigning half of each cross block to either participating group gives
\begin{equation}
  R^{[p]}
  =
  R^{(p,p)}
  +\frac12\sum_{q\ne p}
  \left[
    R^{(p,q)}+R^{(q,p)}
  \right],
  \qquad
  \sum_{p=1}^{P_g}R^{[p]}=R^{\mathrm{coef}} .
  \label{eq:closed-group-accounting}
\end{equation}
Trace and divergence inherit the same identity:
\(k^{[p]}=\tfrac12\operatorname{tr}R^{[p]}\),
\(q^{[p]}=-\nabla_h\cdot R^{[p]}\),
\(\sum_p k^{[p]}=k^{\mathrm{coef}}\), and
\(\sum_p q^{[p]}=q^{\mathrm{coef}}\). This accounting is fixed by \(G\), the
sampled metric \(W\), \(\lambda_{\mathrm{ridge}}\), the group definitions, and
the half-share convention. Exact closure of the sum does not make the named
groups unique physical objects.

Separability is therefore tested in the decoded velocity space. Let
\(M_u\) denote the full-grid kinetic-energy metric and
\(\mathcal S_p=\operatorname{range}(\Phi_p)\) the decoded velocity span of
group \(p\). For bases \(Q_p\) orthonormal in this metric, the pairwise
principal cosines are
\begin{equation}
  Q_p^{\mathsf T}M_uQ_p=\mathbf I,
  \qquad
  \cos\vartheta_j^{(p,q)}
  =
  \sigma_j\!\left(Q_p^{\mathsf T}M_uQ_q\right).
  \label{eq:span-identifiability}
\end{equation}
Values near unity identify directions that the two groups represent almost
interchangeably, even when their exact intersection is trivial. We probe this
overlap with two complementary perturbations: a field-preserving relabelling
that changes the group assignment at fixed total velocity, and a
group-preserving reallocation that retains exact span membership while changing
the total velocity.

For a pair \((p,q)\) and principal-cosine threshold
\(\tau_{\mathrm{PA}}\), let
\(\mathbf a_j,\mathbf b_j\) be the principal vectors of
Eq.~\eqref{eq:span-identifiability} for the \(n_{\mathrm{PA}}\) directions with
\(\cos\vartheta_j\ge\tau_{\mathrm{PA}}\). Their bisectors
\begin{equation}
  \mathbf d_j
  =
  \frac{\mathbf a_j+\mathbf b_j}{\|\mathbf a_j+\mathbf b_j\|_{M_u}},
  \qquad
  \bigl\|\mathbf d_j-\Pi_{\mathcal S_p}\mathbf d_j\bigr\|_{M_u}
  =
  \sin\tfrac{\vartheta_j}{2},
  \label{eq:exchange-basis}
\end{equation}
are mutually orthonormal; \(\sin(\vartheta_j/2)\) is their distance from either
group span. For snapshot \(s\), define
\(\pi_{j,s}^{p}=\langle\mathbf u_{p,s},\mathbf d_j\rangle_{M_u}\) and
\(T_{j,s}=\pi_{j,s}^{p}+\pi_{j,s}^{q}\). A common parameter tuple
\((\tau_{\mathrm{PA}},\alpha_{\mathrm{alloc}},t)\) is applied across the
ensemble, while the exchange coefficients remain snapshot-specific. With
\(\alpha_{\mathrm{alloc}}\in[0,1]\) and \(t\in[0,1]\), the field-preserving
relabelling is
\begin{equation}
  \mathbf u_{p,s}\mapsto\mathbf u_{p,s}
  +t\!\!\sum_{j\le n_{\mathrm{PA}}}\!\!\delta_{j,s}\mathbf d_j,
  \quad
  \mathbf u_{q,s}\mapsto\mathbf u_{q,s}
  -t\!\!\sum_{j\le n_{\mathrm{PA}}}\!\!\delta_{j,s}\mathbf d_j,
  \quad
  \delta_{j,s}=(1-\alpha_{\mathrm{alloc}})T_{j,s}-\pi_{j,s}^{p}.
  \label{eq:field-preserving-relabelling}
\end{equation}
At \(t=0\), the original group fields are recovered. For every \(t\), the
increments cancel snapshot by snapshot, so both \(\widehat U^G\) and
\(R^{\mathrm{coef}}\) remain unchanged. The group-preserving counterpart
projects the increments onto \(\mathcal S_p\) and \(\mathcal S_q\); group
membership is then exact, but the total velocity changes.

The cost is measured in the quantity not preserved. For the field-preserving
case, it is the snapshot-level membership residual
\begin{equation}
  \varrho_{p,s}
  =
  \frac{\bigl\|\widetilde{\mathbf u}_{p,s}
  -\Pi_{\mathcal S_p}\widetilde{\mathbf u}_{p,s}\bigr\|_{M_u}}
       {\|\mathbf u_{p,s}\|_{M_u}},
  \qquad
  \bigl\|\widetilde{\mathbf u}_{p,s}
  -\Pi_{\mathcal S_p}\widetilde{\mathbf u}_{p,s}\bigr\|_{M_u}
  =
  t\Bigl(\textstyle\sum_{j\le n_{\mathrm{PA}}}
  \delta_{j,s}^{2}\sin^{2}\tfrac{\vartheta_j}{2}\Bigr)^{1/2},
  \label{eq:membership-residual}
\end{equation}
where the tilde denotes the relabelled group field. For the group-preserving
case, the corresponding cost is the relative change in \(\widehat U^G\).
At each parameter tuple, we track pair-sign topology, the dominant pair, the
cosine of \(\boldsymbol\eta^R=\{\eta_{pq}^R\}_{p\le q}\), and
\(\mathcal C_R\).

A budgeted stability test also requires a complete parameter grid and a
prescribed reduction of the snapshot- and group-level residuals to an ensemble
gate. Appendix~\ref{app:relabel-protocol} records the protocol and its evidence
status. Stability under the resulting tests is necessary, but not sufficient,
for a representation-relative component reading; it does not establish
causality or uniqueness under another grouping.

\section{Fidelity, conditioning, and covariance stability}
\label{sec:coordinate-verification}

Before interpreting group contributions, we require the frozen representation
to satisfy three distinct conditions: the vorticity fit must remain consistent
with the decoded velocity, directional geometry must avoid nearly duplicate
atoms, and stress accounting must retain complete self--cross covariance. A
small field error does not imply either of the last two.
Table~\ref{tab:fields-ensembles} defines the stress fields and ensemble roles
used from this section onward. The tests below isolate the three requirements
and freeze one geometry; Sec.~\ref{subsec:jhtdb-group-attribution} then tests
whether its group attribution is separable.

\begin{table}
\caption{\label{tab:fields-ensembles}
Stress definitions and HIT ensemble roles.}
\footnotesize
\setlength{\tabcolsep}{4pt}
\begin{ruledtabular}
\begin{tabular}{@{}p{0.16\textwidth}p{0.48\textwidth}p{0.25\textwidth}@{}}
Field & Reynolds stress & Purpose\\
\colrule
DNS
  & \(R^{\mathrm{DNS}}=\operatorname{Cov}_s(\mathbf u'_s)\)
  & reference\\
\multicolumn{1}{l}{Gaussian\footnote{Only \(R^{\mathrm{coef}}\) is
decomposed by group.}}
  & \(R^{\mathrm{coef}}=\Phi C_A\Phi^{\mathsf T}\)
  & group decomposition\\
Gaussian + mean
  & \(R^{\mathrm{dep}}=\operatorname{Cov}_s(\widehat{\mathbf u}_s)\)
  & fidelity\\
\end{tabular}
\end{ruledtabular}

\vspace{5pt}
\footnotesize
\setlength{\tabcolsep}{3pt}
\begin{ruledtabular}
\begin{tabular}{llll}
Ensemble & \(N_o\!\times\!N_t=S\) & Sampling relation\footnote{
\(N_o\), \(N_t\), and \(S\) denote origins, time indices, and snapshots.
``Shared times'' means reused time indices; all origin--time pairs remain
disjoint. Pooled and Aggregate contain Selection data and are therefore not
selection-free evaluations.} & Role\\
\colrule
Geometry
& \(31\times8=248\)
& base origins; canonical 8 times
& geometry\\
Selection
& \(16\times8=128\)
& new origins; shared 8 times
& parameters and groups\\
Reserved
& \(16\times8=128\)
& new origins; shared 8 times
& selection-free transfer\\
Pooled
& \(32\times8=256\)
& Selection \(\cup\) Reserved
& recentered aggregate\\
T202
& \(1\times202=202\)
& new origin; 199 new + 3 shared times
& selection-free time window\\
Aggregate
& \(63\times8=504\)
& Geometry \(\cup\) Selection \(\cup\) Reserved
& sampling check\\
\end{tabular}
\end{ruledtabular}
\end{table}

\subsection{Graph-norm fidelity and Gram conditioning}
\label{subsec:verify-coordinate}

Under Biot--Savart recovery, a solenoidal vorticity mode at
\(\mathbf k\) contributes to velocity with weight \(|\mathbf k|^{-1}\);
pure vorticity \(L^2\) therefore emphasizes high-wavenumber error more strongly
than recovered-velocity fidelity. The graph norm combines the induced
\(|\mathbf k|^{-2}\) velocity metric with a controlled vorticity term, retaining
sensitivity to both velocity-carrying and derivative-bearing content.

At the retained geometry, replacing vorticity-\(L^2\) fitting by the graph norm
reduces decoded-velocity error from \(0.326\) to \(0.239\) and stress error from
\(0.127\) to \(0.105\) (Table~\ref{tab:coord-verify}). Because this paired
comparison changes only the fitting metric, the subsequent analysis uses
graph-norm amplitudes.

Directional geometry introduces a second requirement. Isotropic envelopes
impose one local length scale, whereas
\[
\Sigma_i
=
Q_i\operatorname{diag}(\sigma_{i,1}^2,\sigma_{i,2}^2,\sigma_{i,3}^2)
Q_i^{\mathsf T},
\]
permits direction-dependent widths and orientation. This freedom can also
produce nearly collinear curl-lifted atoms where envelopes overlap. The
thresholded penalty \(\mathcal R_{\mathrm{Gram}}\) in
Eq.~\eqref{eq:gram-regularization} acts on the graph-metric Gram
\(K_G=G^{\mathsf T}WG\), limiting this degeneracy without enforcing global
orthogonality.

The post-hoc rows in Table~\ref{tab:coord-verify} isolate fixed-geometry
sensitivity: removing orientation or isotropizing the inferred widths worsens
both field error and Gram conditioning. They do not rank independently learned
geometry families.

\begin{table*}
\caption{\label{tab:coord-verify}
Matched field fidelity and Gram conditioning.}
\footnotesize
\setlength{\tabcolsep}{4.0pt}
\begin{ruledtabular}
\begin{tabular}{llccccccc}
& &
\multicolumn{4}{c}{Full-field relative errors\footnote{
\(e_u=\|\widehat U-U\|_F/\|U\|_F\), while \(e_{\rm graph}\) and \(e_R\)
are the corresponding graph-norm and Reynolds-stress errors. Evaluation restores the exact
per-snapshot constant channel and uses
\(R^{\rm dep}=\operatorname{Cov}(\widehat{\mathbf u}_s)\), with
\(\widehat{\mathbf u}_s=\overline{\mathbf u}_{s,h}+\Phi a_s\), against
\(R^{\rm DNS}=\operatorname{Cov}(\mathbf u')\).
\(e_R^{\rm T202}\) is the same full represented-stress error on the
\(S=202\) temporal ensemble.}}
&
\multicolumn{3}{c}{Graph-metric Gram\footnote{
\(K_G=G^{\mathsf T}WG\) uses unit-normalized columns of the sampled
graph-norm design. Here \(\kappa\) and \(r_{\rm eff}\) denote its condition
number and effective rank, and
\(E_{\rm coh}=\operatorname*{mean}_{m<n}
[\max(0,|(C_G)_{mn}|-\tau_{\mathrm{Gram}})]^2\), with
\(\tau_{\mathrm{Gram}}=0.5\).}}\\
Geometry\footnote{
All rows use \(N_G=4096\), \(M_b=12288\), the retained
\((N_{\rm geom},N_{\rm opt})=(64,40)\) state, and the Pooled ensemble of
Table~\ref{tab:fields-ensembles}; selection-free results are in
Table~\ref{tab:jhtdb-transfer}. Isotropized and axis-aligned rows are post-hoc
transformations of the same geometry. Isotropized replaces each width triplet
by its geometric mean and sets \(Q_i=I\); axis-aligned sets only \(Q_i=I\).
Fixed isotropic uses \(\sigma=5\Delta x\). Boldface marks the best comparable
value, maximizing only \(r_{\rm eff}\).}
&
Fit\footnote{
Graph fitting uses \(\ell=0.25\Delta x\) and relative ridge \(10^{-8}\).
The vorticity-\(L^2\) row uses \(W=I\); its Gram diagnostics are not
metric-comparable and are omitted.}
& \(e_u\)
& \(e_{\rm graph}\)
& \(e_R\)
& \(e_R^{\rm T202}\)
& \(\kappa(K_G)\)
& \(r_{\rm eff}(K_G)\)
& \(E_{\rm coh}\)\\
\colrule
fixed iso. & graph
& \(0.280\) & \(0.390\) & \(0.166\) & \(0.233\)
& \(6.6\times10^{19}\) & \(427\) & \(1.3\times10^{-4}\) \\
post-hoc isotropized & graph
& \(0.255\) & \(0.359\) & \(0.139\) & \(0.199\)
& \(3.3\times10^{12}\) & \(1374\) & \(2.6\times10^{-5}\) \\
post-hoc axis-aligned & graph
& \(0.244\) & \(0.343\) & \(0.112\) & \(0.159\)
& \(1.3\times10^{8}\) & \(1815\) & \(8.2\times10^{-6}\) \\
oriented (retained) & graph
& \(\mathbf{0.239}\) & \(\mathbf{0.337}\)
& \(\mathbf{0.105}\) & \(\mathbf{0.150}\)
& \(\mathbf{1.0\times10^{7}}\) & \(\mathbf{1999}\)
& \(\mathbf{2.2\times10^{-6}}\) \\
oriented (retained) & vort. \(L^2\)
& \(0.326\) & \(0.442\) & \(0.127\) & \(0.181\)
& -- & -- & -- \\
\end{tabular}
\end{ruledtabular}
\end{table*}

Independent geometry determination resolves the benefit of direction-dependent
widths, but not a further gain from free orientation under the retained
coherence penalty. Removing that penalty lowers field error while degrading
condition number and effective rank. We therefore retain the richer oriented,
regularized geometry without claiming that orientation is separately resolved.
Appendix~\ref{app:geometry-family-evidence} gives the matched comparison and
uncertainty.

Global references bound this choice rather than determine it. Fourier is
cheaper and gives comparable stress fidelity in the seam-free matched test;
the frozen POD comparison is limited by snapshot rank and matched storage.
The Gaussian representation is retained for its localized, groupable geometry,
not as a universal fidelity optimum. The comparisons are reported in
Sec.~\ref{subsec:jhtdb-transfer} and Appendix~\ref{app:reproducibility}.

\subsection{Self--cross covariance accounting}
\label{subsec:verify-attribution}

For a fixed linear representation, Reynolds stress depends on the complete
amplitude covariance. Equation~\eqref{eq:covariance-lift} reproduces the
stress of the represented Gaussian velocity exactly; its discrepancy from the
DNS sample stress is therefore representation error. Once the frozen atoms are
partitioned, Eqs.~\eqref{eq:group-pair-accounting} and
\eqref{eq:closed-group-accounting} separate within-group self covariance from
between-group cross covariance.

We examine the fixed shape, scale, aspect-ratio, and lift-strength partitions
specified in Appendix~\ref{app:reproducibility}. Each envelope label is
inherited by its three curl-lifted atoms, and half of every cross block is
assigned to each participating group through the closed field \(R^{[p]}\),
preserving the total represented stress and its linear trace and divergence
diagnostics.

\begin{table}
\caption{\label{tab:coord-closure}
Cross-group covariance accounting on the Pooled ensemble.}
\footnotesize
\setlength{\tabcolsep}{5.0pt}
\begin{ruledtabular}
\begin{tabular}{llccccc}
Geometry & Fit & accounting residual (max)\footnote{The Pooled ensemble has
\(S=256\) (Table~\ref{tab:fields-ensembles}). The self-only deficit is
\(\|\sum_pR^{(p,p)}-R^{\rm coef}\|_F/\|R^{\rm coef}\|_F\); retaining
symmetrized cross blocks gives \(\sum_pR^{[p]}=R^{\rm coef}\) to the reported
residual. Daggered entries are degenerate one-group partitions and are excluded
from comparisons.} & shape & scale & aspect
& lift\footnote{
Shape, scale, aspect, and lift use the fixed descriptive partitions recorded
in Appendix~\ref{app:reproducibility}. Their cutoffs are representation labels,
not universal physical thresholds.}\\
\colrule
fixed iso. & graph & \(\le2.7\times10^{-12}\)
& \(0.00^{\dagger}\) & \(0.00^{\dagger}\) & \(0.00^{\dagger}\)
& \(\mathbf{10.95}\) \\
post-hoc isotropized & graph & \(\le2.7\times10^{-12}\)
& \(0.00^{\dagger}\) & \(4.1\times10^{3}\) & \(0.00^{\dagger}\)
& \(2.6\times10^{4}\) \\
post-hoc axis-aligned & graph & \(\le2.7\times10^{-12}\)
& \(3.4\times10^{2}\) & \(36.14\) & \(4.8\times10^{2}\)
& \(1.2\times10^{2}\) \\
oriented (retained) & graph & \(\le2.7\times10^{-12}\)
& \(47.21\) & \(14.64\) & \(60.60\) & \(23.39\) \\
oriented (retained) & vort. \(L^2\) & \(\le2.7\times10^{-12}\)
& \(62.22\) & \(20.14\) & \(74.17\) & \(30.48\) \\
\end{tabular}
\end{ruledtabular}
\end{table}

Cross-group covariance is load-bearing for every nondegenerate partition in
Table~\ref{tab:coord-closure}: omitting it gives self-only deficits from
\(14.64\) to \(60.60\). Restoring the symmetrized cross blocks reproduces
\(R^{\mathrm{coef}}\) to a relative residual below
\(2.7\times10^{-12}\), verifying the exact attribution numerically.

The represented stress is therefore a balance of self and cross covariance,
not a sum of independent group stresses. All subsequent group analyses use
\(R^{[p]}\), not \(R^{(p,p)}\). Exact accounting completes the tensor
decomposition but does not make the allocation unique; span overlap is tested
in Sec.~\ref{subsec:jhtdb-group-attribution}.

Across non-sparse shape partitions, the self-positive/cross-negative pattern
and dominant sheet--tube cross block persist. Their magnitudes, and the
dominant lift pair, remain threshold-sensitive. The later physical reading is
therefore explicitly grouping-relative.

\subsection{Selection of the frozen geometry}
\label{subsec:verify-geometry-selection}

A useful frozen representation must resolve stress without reorganizing its
signed group covariance under modest refinement. We define this empirical
covariance stability by origin-block repeatability and agreement between nearby
geometry states; it is a selection criterion, not a claim of group
identifiability.

We compare the prescribed grid
\[
N_{\mathrm{geom}}\in\{15,64,128\},
\qquad
N_{\mathrm{opt}}\in\{12,40,80,160\},
\]
and use \(B=N_{\mathrm{geom}}N_{\mathrm{opt}}\) as an exposure proxy. Every
state is evaluated on the same Selection ensemble, whose origins and
origin--time pairs are disjoint from Geometry although its eight time indices
are shared. Geometry is frozen and only amplitudes are refitted.

\begin{figure*}
  \centering
  \includegraphics[width=\textwidth]
  {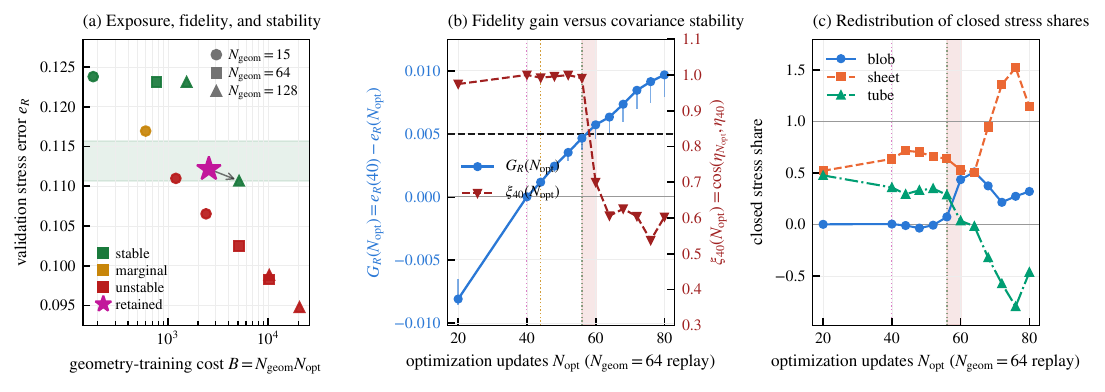}
  \caption{
  Geometry selection by stress fidelity and covariance stability.
  (a) Selection stress error \(e_R\) versus exposure
  \(B=N_{\mathrm{geom}}N_{\mathrm{opt}}\); marker shape gives
  \(N_{\mathrm{geom}}\), colour gives stability status, and the green region
  marks the \(\epsilon_R=0.005\) fidelity band.
  (b) In the \(N_{\mathrm{geom}}=64\) replay, the first resolved fidelity gain
  coincides with the closed-share cosine falling from \(0.991\) to \(0.700\).
  (c) The same transition redistributes the signed group shares while leaving
  the decoded span nearly unchanged.
  }
  \label{fig:geometry-budget-selection}
\end{figure*}

No candidate satisfies the original joint gate on fidelity, Gram geometry,
accounting, and covariance stability. We therefore apply the declared
secondary rule. Let \(\mathcal S_{\mathrm{cov}}\) be the candidates that pass
the sign, origin-block, and cross-state stability checks, and retain
\[
  \mathcal B_{\mathrm{stab}}
  =
  \left\{
  c\in\mathcal S_{\mathrm{cov}}:
  e_R(c)
  \le
  \min_{d\in\mathcal S_{\mathrm{cov}}}e_R(d)
  +\epsilon_R
  \right\},
  \qquad
  \epsilon_R=0.005.
\]
The band in Fig.~\ref{fig:geometry-budget-selection}(a) is
\(\mathcal B_{\mathrm{stab}}\). Complete gates and resampling rules are given in
Appendix~\ref{app:reproducibility}. Candidate-level reallocation sweeps are not
available for every member of this set, so covariance stability is not treated
as a shared identifiability result.

The geometry-data axis is resolved at \(N_{\mathrm{opt}}=40\): increasing
\(N_{\mathrm{geom}}\) from \(64\) to \(128\) improves \(e_R\) by only
\(0.0015\), with paired \(95\%\) interval \(0.0010\)--\(0.0020\), below
\(\epsilon_R\) [Fig.~\ref{fig:geometry-budget-selection}(a)].

The optimization axis is limited instead by covariance stability. In the dense
\(N_{\mathrm{geom}}=64\) replay, let \(\xi_{40}\) be the cosine between the
signed blob--sheet--tube share vector and its value at
\(N_{\mathrm{opt}}=40\). The last stable point, \(N_{\mathrm{opt}}=56\), has
\(\xi_{40}=0.991\) and an unresolved fidelity gain of \(0.0046\). At
\(N_{\mathrm{opt}}=60\), the gain reaches \(0.0057\) while \(\xi_{40}\) falls
to \(0.700\) [Fig.~\ref{fig:geometry-budget-selection}(b)]. Across this
transition, the decoded-span overlap remains at least \(0.98\), but the Gram
condition number rises from approximately \(2.5\times10^7\) to
\(1.0\times10^8\) and the signed shares reorganize
[Fig.~\ref{fig:geometry-budget-selection}(c)]. Lower stress error is therefore
obtained through more dependent directions and stronger covariance
redistribution, not a materially different decoded span.

We consequently freeze
\((N_{\mathrm{geom}},N_{\mathrm{opt}})=(64,40)\), the lowest-exposure member of
\(\mathcal B_{\mathrm{stab}}\). The selected state resolves stress fidelity
within the tested budget while retaining the empirical covariance pattern.
Whether its named groups are separable remains a distinct physical question,
addressed in Sec.~\ref{subsec:jhtdb-group-attribution}.

\section{Reynolds-stress structure in homogeneous isotropic turbulence}
\label{sec:jhtdb-attribution}
Homogeneity removes mean-shear and wall-normal organization, making HIT a
controlled setting in which to separate stress fidelity from the covariance
structure introduced by the representation. We freeze the selected geometry
and refit only amplitudes on ensembles with distinct evidence roles
(Table~\ref{tab:fields-ensembles}). Full-field fidelity uses
\(R^{\mathrm{dep}}\), including the exact constant-velocity channel; component
analysis partitions only the Gaussian-channel stress
\(R^{\mathrm{coef}}\). Sampling, resolved bandwidth, preprocessing, and data
splits are specified in Appendix~\ref{app:reproducibility}.

\subsection{Transfer of the frozen representation}
\label{subsec:jhtdb-transfer}

The frozen geometry transfers across origin-disjoint samples within the same
HIT regime. On the 16 reserved origins, \(e_R=0.113\), compared with
\(0.112\) on the selection origins; the energy ratio and median velocity error
also agree (Table~\ref{tab:jhtdb-transfer}). Because these roles reuse the same
eight time indices, their agreement establishes spatial reuse, not temporal
independence.

Boundary treatment and finite-window sampling set the absolute error level.
Trimming four cells per face reduces reserved-origin \(e_R\) from \(0.113\) to
\(0.073\). T202 changes both origin and time window and gives
interior/full-grid \(e_R=0.095/0.150\), \(\chi_k=0.926\), and
\(e_u^{\mathrm{med}}=0.329\). It is therefore a correlated finite-window
diagnostic rather than a measure of temporal degradation. The seam-free
whole-box comparison gives matched-budget errors of \(0.203\) (Gaussian) and
\(0.216\) (Fourier), but at a coarser stride and only within that ensemble.

Two stress references must remain distinct. Full-field fidelity uses
\(R^{\mathrm{dep}}\), including the exact constant channel, whereas the group
analysis partitions \(R^{\mathrm{coef}}\), whose DNS-referenced error is
\(0.436\)--\(0.499\). The former measures what the complete representation
reconstructs; the latter states the fidelity of the tensor whose internal
covariance balance is analyzed.

\begin{table}
\caption{\label{tab:jhtdb-transfer}
HIT transfer and sampling consistency.}
\footnotesize
\setlength{\tabcolsep}{4pt}
\begin{ruledtabular}
\begin{tabular}{lccccc}
Role\footnote{Roles and sample counts are defined in
Table~\ref{tab:fields-ensembles}. Reserved is selection-free; Pooled contains
Selection; Aggregate also contains Geometry; T202 is one correlated finite
window. Pooled and Aggregate are recentered sampling checks, not additional
held-out transfer tests.}
& \(S\) & \(e_u^{\rm med}\)
& \(e_R\)\footnote{Each row is centered independently and compared with its
own \(R^{\rm DNS}\). Here \(e_R\) evaluates the full represented stress
\(R^{\rm dep}\), whereas \(e_R^{\rm coef}\) evaluates the Gaussian-channel
stress used for group attribution. Trim-four \(e_R\) is \(0.073\) for Reserved
and Selection and \(0.095\) for T202.}
& \(e_R^{\rm coef}\) & \(\chi_k\)\\
\colrule
Reserved
& \(128\) & \(0.241\) & \(\mathbf{0.113}\) & \(0.499\) & \(0.958\)\\
Selection
& \(128\) & \(0.245\) & \(0.112\) & \(0.498\) & \(0.957\)\\
Pooled
& \(256\) & \(0.244\) & \(0.105\) & \(0.482\) & \(0.958\)\\
T202
& \(202\) & \(0.329\) & \(0.150\) & \(0.436\) & \(0.926\)\\
Aggregate
& \(504\) & \(0.241\) & \(0.097\) & \(0.493\) & \(0.959\)\\
\end{tabular}
\end{ruledtabular}
\end{table}

\begin{figure*}
\begingroup
  \centering
  \includegraphics[width=\textwidth]
  {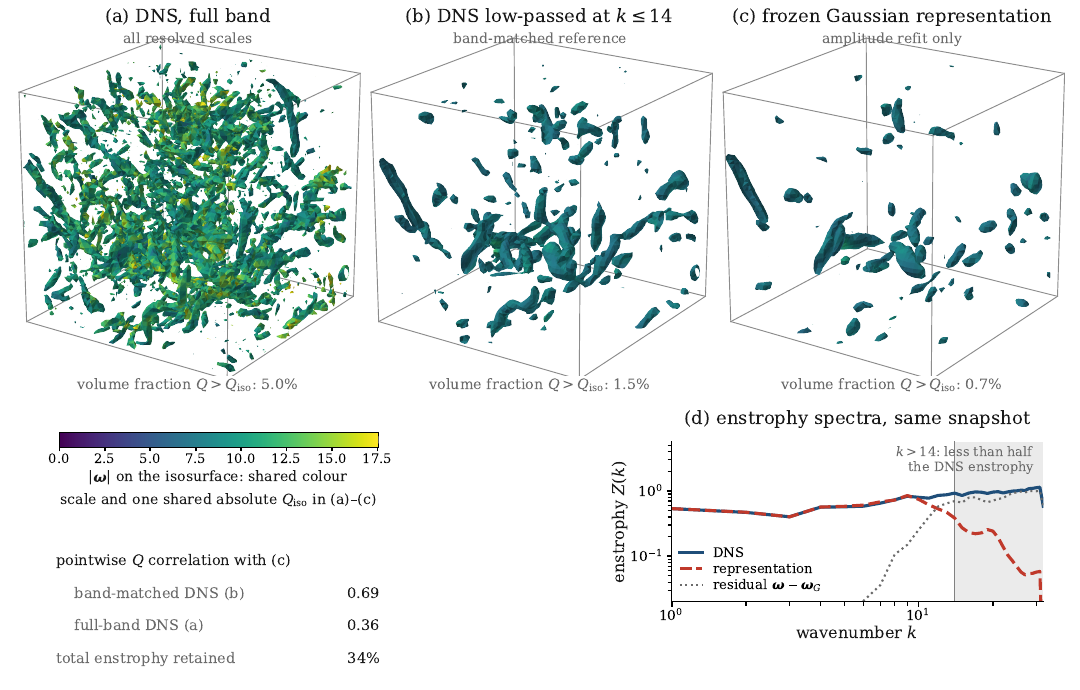}
  \caption{\label{fig:vortex-structure}
  Resolved vortical organization on a median-fidelity reserved snapshot.
  (a) DNS, (b) DNS low-pass filtered at \(k=14\), and (c) the frozen
  representation. All panels use the DNS 95th-percentile \(Q\) threshold and
  a common vorticity scale. The \(64^3\) reference has
  \(\Delta x\approx8.5\,\eta_K\), so the comparison is between resolved
  bandwidths.}
\endgroup
\end{figure*}

The frozen geometry also preserves the spatial organization of the dominant
band-limited vortical motions [Fig.~\ref{fig:vortex-structure}]. On the
median-fidelity reserved snapshot,
the represented \(Q\) field correlates more strongly with the DNS low-pass
filtered at \(k=14\) than with the full resolved field (\(0.69\) against
\(0.36\)), while retaining \(34.3\%\) of the DNS enstrophy. At the same absolute
\(Q\) threshold, the represented structures occupy \(0.75\%\) of the interior,
compared with \(1.50\%\) for the band-matched DNS and \(5.00\%\) for the full
field. The representation therefore recovers the dominant band-limited
vortical regions while smoothing finer-scale and weaker threshold-exceeding
structures. Because the reference is itself sampled at
\(\Delta x\approx8.5\,\eta_K\), the comparison is between two resolved
bandwidths, not against the dissipative range.

The frozen geometry thus transfers its resolved vortical organization
across origins and a changed finite window. The next test asks whether the
resulting covariance also retains the local tensorial state of Reynolds stress.

\subsection{Tensorial stress fidelity}
\label{subsec:jhtdb-stress-state}

The frozen vorticity representation preserves the local tensorial organization
of Reynolds stress beyond its scalar turbulent kinetic energy (TKE). Neither \(k\)
nor \(R\) is fitted directly; both follow from the covariance of the decoded
velocity ensemble. The decomposition
\(R=2k(b+\mathbf I/3)\), with
\(k=\tfrac12\operatorname{tr}R\) and
\(b=R/(2k)-\mathbf I/3\), sets the three nested levels of the test: \(k\)
measures local fluctuation intensity, whereas the eigenvalues and eigenvectors
of \(b\) determine componentality and principal directions.

\begin{figure}
  \centering
  \begin{overpic}[width=\linewidth]
  {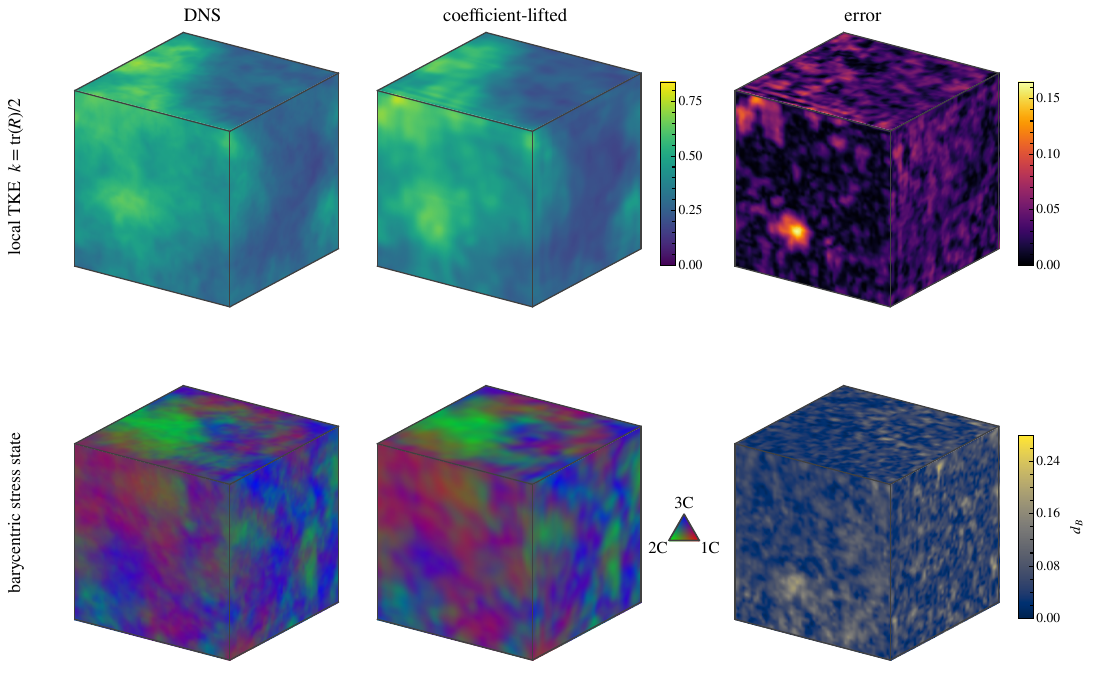}
  \put(37.0,60.2){\colorbox{white}{\makebox[0.20\linewidth][c]{\scriptsize represented field}}}
  \end{overpic}
  \caption{
  TKE and barycentric stress state on T202 (\(S=202\), \(64^3\)). The top row
  compares DNS and represented TKE; the bottom row compares their barycentric
  states. The errors are \(e_k=0.0755\), median \(d_B=0.045\), and
  \(p_{90}(d_B)=0.085\). Principal-direction fidelity is reported in
  Appendix~\ref{app:stress-orientation}.}
  \label{fig:jhtdb-stress-attribution}
\end{figure}

At the scalar level, the represented TKE field follows the DNS spatial
distribution on T202 with
\[
  e_k=
  \frac{\|k^{\mathrm{dep}}-k^{\mathrm{DNS}}\|_2}
       {\|k^{\mathrm{DNS}}\|_2}
  =0.0755
\]
[Fig.~\ref{fig:jhtdb-stress-attribution}(a--c)]. Because
\(2k=\sum_i\langle u_i'^2\rangle\), this agreement shows that the refitted
amplitudes recover where fluctuation variance is concentrated.

Beyond scalar energy content, the barycentric distance
\(d_B=\|\mathbf y_B^{\mathrm{dep}}-\mathbf y_B^{\mathrm{DNS}}\|_2\)
has median \(0.045\) and 90th percentile \(0.085\)
[Fig.~\ref{fig:jhtdb-stress-attribution}(d--f)]. Because barycentric
coordinates depend on the eigenvalues of \(b\), this agreement resolves local
one-, two-, and three-component states rather than merely matching TKE.

Componentality alone does not determine stress orientation. Where the leading
eigenvector is well conditioned, its angular error has median
\(3.50^\circ\) and 90th percentile \(9.47^\circ\) over \(98.75\%\) of the
interior; the corresponding deviatoric-tensor angles are \(6.83^\circ\) and
\(13.85^\circ\), with the eigengap mask and its sensitivity reported in
Appendix~\ref{app:stress-orientation}. Without exhausting the full-field
discrepancy \(e_R=0.150\), these diagnostics show that the representation
retains the dominant magnitude, componentality, and orientation of the
finite-window stress.

Tensorial fidelity makes the represented stress physically
interpretable as a field. It does not yet show that this field can be assigned
uniquely to the frozen morphological groups.

\subsection{Shared spans and covariance cancellation}
\label{subsec:jhtdb-group-attribution}

The blob-, sheet-, and tube-like groups are morphologically distinct but
kinematically nonseparable within the frozen representation. Their decoded
velocity spans overlap, and the represented stress depends on covariance
between them. Each envelope passes its shape label to its three curl-lifted
columns; the exact constant channel remains unlabelled. Equations~\eqref{eq:group-pair-accounting}
and \eqref{eq:pair-alignment} split the Gaussian-channel stress into signed
self and cross blocks, while \(\mathcal C_R\) measures the cancellation hidden
by their unit net alignment.

\begin{figure}
  \centering
  \includegraphics[width=\linewidth]
  {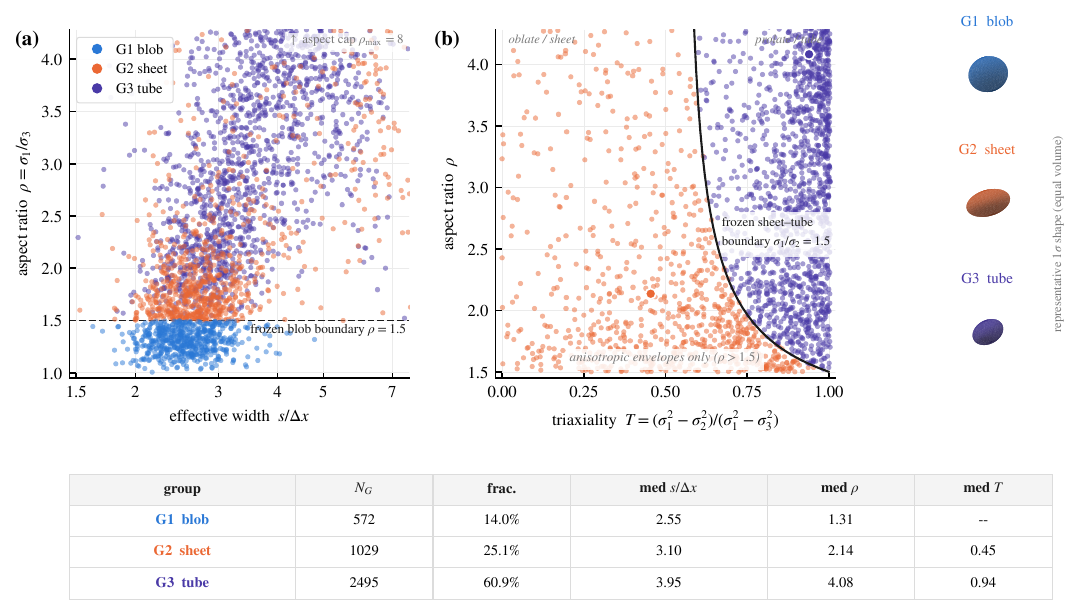}
  \caption{
  Frozen Gaussian-envelope geometry. (a) Effective width and aspect ratio.
  (b) Triaxiality separating oblate sheet-like from prolate tube-like support.
  The blob, sheet, and tube groups contain 572, 1029, and 2495 envelopes;
  \(4.8\%\) reach \(\rho_{\max}=8\).
  }
  \label{fig:jhtdb-group-geometry}
\end{figure}

The labels describe anisotropic support, not separate scales or detected vortex
topologies [Fig.~\ref{fig:jhtdb-group-geometry}]. Effective widths overlap,
whereas aspect ratio and triaxiality distinguish the three groups; \(86.0\%\)
of the envelopes are anisotropic and \(4.8\%\) reach the imposed aspect-ratio
bound. The partition is geometrically informative without implying kinematic
independence.

The span geometry locates the nonadditivity. At
\(\tau_{\mathrm{PA}}=0.95\), 1703 directions---\(55\%\) of the sheet
span---are near-shared with the tube span. Shared--shared products carry
\(99.6\%\) of \(\mathcal C_R\), whereas mutually unrepresentable sheet and
tube content carries \(0.3\%\). The cancellation is therefore organized in
velocity directions that the two labels can both represent, rather than in
content exclusive to either group.

\begin{figure}
  \centering
  \includegraphics[width=\linewidth]
  {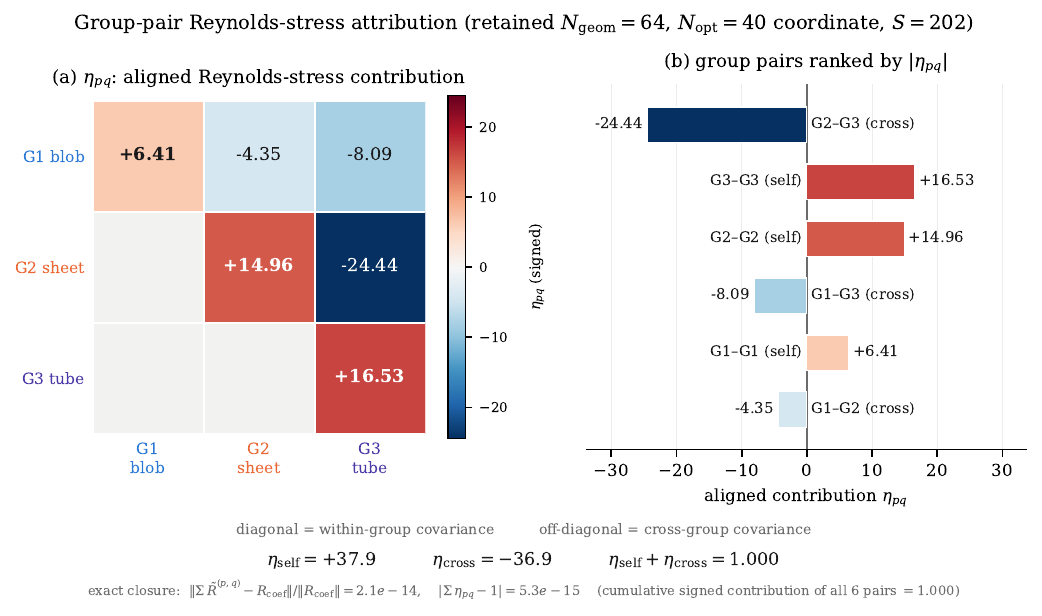}
  \caption{
  Signed group-pair stress alignment on T202. (a) The six self and cross
  coefficients. (b) Their signed magnitude ranking. Self and cross totals are
  \(+37.9\) and \(-36.9\), and sheet--tube is the dominant opposing block.
  }
  \label{fig:jhtdb-pair-attribution}
\end{figure}

The same overlap appears in a strongly nonadditive covariance balance on T202
[Fig.~\ref{fig:jhtdb-pair-attribution}]. The self blocks contribute \(+37.9\)
and the cross blocks \(-36.9\), leaving unit net alignment with
\(\mathcal C_R=74.8\). Every cross block opposes the represented stress;
sheet--tube is the strongest at \(\eta_{23}^{R}=-24.44\), while tube--tube and
sheet--sheet are the largest positive blocks. The tensor accounting closes to
\(2.1\times10^{-14}\). Thus the same labelled groups supply both variance and
the covariance that counterbalances it.

Positive self alignment alone is algebraic, not a turbulence finding. A self
sum above unity necessarily gives a negative total cross sum; Appendix
\ref{app:covariance-sign} provides the derivation. What is specific to this
representation and ensemble is the magnitude of the balance, the opposition of
all three cross blocks, the dominant pair, and their spatial organization.

\begin{figure}
  \centering
  \includegraphics[width=\linewidth]
  {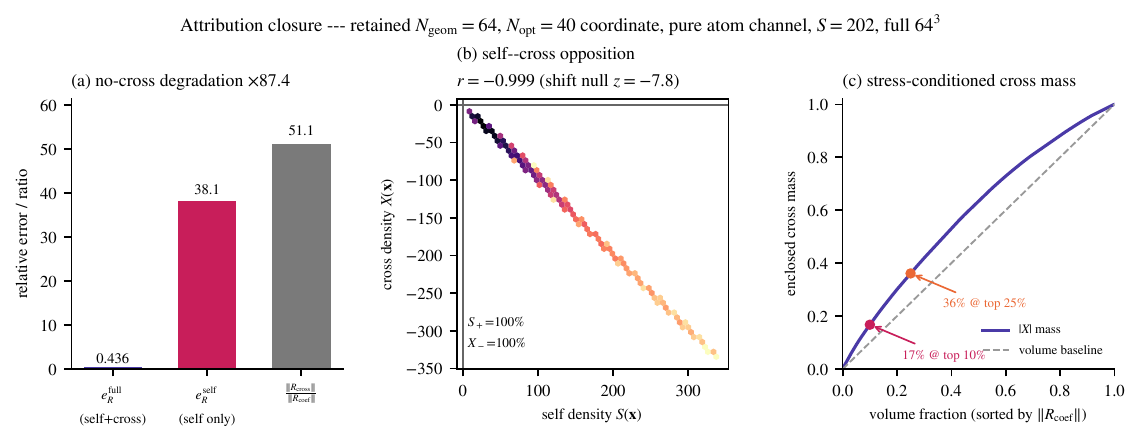}
  \caption{
  Covariance cancellation on T202. (a) Stress-error increase after removing
  cross-group covariance. (b) Pointwise opposition of self and cross densities
  (\(r_{\mathrm{SC}}=-0.9993\)). (c) Cross-covariance mass contained in the
  highest-stress volume fractions.
  }
  \label{fig:jhtdb-covariance-closure}
\end{figure}

Cross covariance is load-bearing, not a small correction to the self blocks
[Fig.~\ref{fig:jhtdb-covariance-closure}]. Its field norm is approximately 51
times that of the assembled Gaussian-channel stress, and omitting it raises the
DNS-referenced stress error by a factor of 87 on T202 and 53.6 on the selection
ensemble. These factors refer to the Gaussian channel being partitioned, whose
own DNS-referenced error is \(0.436\)--\(0.499\).

The counterbalance also occurs in the same physical regions. Summed self and
cross alignment densities oppose almost pointwise
(\(r_{\mathrm{SC}}=-0.9993\)); a periodic-shift null gives \(z=-7.8\) and
\(p_{\mathrm{emp}}<0.001\). The highest-stress \(10\%\) and \(25\%\) of the
volume contain \(17\%\) and \(36\%\) of the cross-covariance mass. Cross
covariance is therefore moderately enriched in intense regions but remains a
distributed counterfield throughout the represented flow.

The controls isolate two requirements for this balance: nonorthogonal group
geometry and correlated amplitudes (Table~\ref{tab:jhtdb-grouping-controls}).
Orthogonal Fourier or POD regrouping of the same decoded field gives
\(\mathcal C_R=1.000\). A shell-matched random field decoded through the frozen
dictionary retains strong cancellation (\(70.4\pm1.4\)), whereas row-shuffling
the amplitudes removes inter-atom correlation and returns
\(1.001\pm0.003\). The cancellation therefore belongs to the
dictionary--covariance pair, not to the represented field or marginal
amplitude statistics alone.

\begin{table}
\caption{\label{tab:jhtdb-grouping-controls}
Cancellation controls on the selection ensemble.}
\footnotesize
\setlength{\tabcolsep}{4pt}
\begin{ruledtabular}
\begin{tabular}{lccc}
Control\footnote{The ensemble has \(S=128\). The first block compares the shape partition with orthogonal
regroupings of the same decoded field. The second holds the shape grouping
fixed while changing amplitude statistics; stochastic rows report
mean \(\pm\) standard deviation over 24 seeds. The matched-covariance row is a
consistency check, not a null model.}
& \(\kappa\)\footnote{Condition number of the grouping-basis Gram matrix.}
& \(\mathcal C_R\)
& Drop-cross\footnote{Stress-error inflation after discarding cross blocks.
This table reports the selection-ensemble value 53.6 for the shape partition;
the corresponding T202 value in the text is 87.}\\
\colrule
\multicolumn{4}{l}{\emph{Shape partition and orthogonal regroupings}}\\
Shape partition & \(1.7\times10^{3}\) & \(83.16\) & \(53.6\)\\
Fourier, \(\rho=0.25,0.50\) & \(1.000\) & \(1.000\) & \(1.02\)\\
Fourier, energy matched & \(1.000\) & \(1.000\) & \(0.99\)\\
Fourier, equal energy & \(1.000\) & \(1.000\) & \(0.99\)\\
POD, energy matched & \(1.000\) & \(1.000\) & \(1.00\)\\
POD, equal energy & \(1.000\) & \(1.000\) & \(1.00\)\\
\colrule
\multicolumn{4}{l}{\emph{Null amplitudes through the frozen dictionary}}\\
Random field, shell matched & --- & \(70.4\pm1.4\) & \(41.3\pm4.0\)\\
Row-shuffled amplitudes & --- & \(1.001\pm0.003\) & \(0.99\)\\
Random field, per-mode matched & --- & \(82.9\pm2.1\) & \(49.4\pm5.0\)\\
\colrule
\multicolumn{4}{l}{\emph{Covariance consistency}}\\
Matched-covariance Gaussian & --- & \(80.5\pm1.6\) & \(53.6\pm2.3\)\\
\end{tabular}
\end{ruledtabular}
\end{table}

Gaussian shape is not required for the effect. A fixed-field control replaces
the orthonormal POD coordinate by square nonorthogonal coordinates in the same
POD span. At fixed \(e_R^{\mathrm{coef}}=0.498\), increasing the learned
atom-Gram condition number from 38 to \(5.3\times10^3\) raises
\(\mathcal C_R\) from 1.00 to 13.80 and the drop-cross factor from 1.06 to
7.37 (Table~\ref{tab:second-representation}). A sparse solve on the same atoms
reduces cancellation while leaving half the field unresolved. Span geometry
therefore creates the opportunity for redistribution, and amplitude allocation
determines how strongly it is used.

A 29-coordinate sweep further shows that condition number alone does not locate
the transition: the first near-shared direction appears only after
\(\kappa\) exceeds \(2\times10^2\), while one more ill-conditioned coordinate
has no principal cosine above \(0.95\). Conditioning, span overlap, and
covariance cancellation are complementary rather than interchangeable
diagnostics.

\begin{table}
\caption{\label{tab:second-representation}
Coordinate control in a fixed POD span.}
\footnotesize
\setlength{\tabcolsep}{4pt}
\begin{ruledtabular}
\begin{tabular}{lcccc}
Representation\footnote{All rows use the selection ensemble (\(S=128\)).
The POD modes are orthonormal. The ridge rows are
square, nonorthogonal coordinates learned in the same selection-ensemble POD
span and grouped by spectral content; they hold
\(e_R^{\rm coef}=0.498\) fixed. Hence \(\kappa\) and \(\mu_{\max}\) describe
the learned atom Gram, not the POD Gram. Orthogonal matching pursuit (OMP) changes the amplitude solve on
identical atoms; Frozen Gaussian is an external reference.}
& \(\kappa\) & \(\mu_{\max}\) & \(\mathcal C_R\) & Drop-cross\\
\colrule
\multicolumn{5}{l}{\emph{Nonorthogonal coordinates, ridge amplitudes}}\\
\(\alpha=5.0\) & \(38\) & \(0.382\) & \(1.000\)
  & \(1.06\)\\
\(\alpha=1.0\) & \(1.7\times10^{3}\) & \(0.796\)
  & \(6.62\) & \(2.93\)\\
\(\alpha=0.2\) & \(4.7\times10^{3}\) & \(0.837\)
  & \(12.65\) & \(6.65\)\\
\(\alpha=0.05\) & \(5.3\times10^{3}\) & \(0.841\)
  & \(13.80\) & \(7.37\)\\
\colrule
\multicolumn{5}{l}{\emph{Greedy sparse fit, identical atoms}}\\
OMP, \(16\) nonzeros & \(5.3\times10^{3}\) & \(0.841\)
  & \(2.39\) & \(0.74\)\\
\colrule
Frozen Gaussian & \(1.0\times10^{7}\) & \(0.953\)
  & \(83.16\) & \(53.65\)\\
\end{tabular}
\end{ruledtabular}
\end{table}

The signed balance is reproducible, although its magnitude depends on the
partition and finite sample. Every tested time block and 11 spatial controls
retain positive self blocks, negative cross blocks, and sheet--tube as the
strongest opposing pair; \(\mathcal C_R\) ranges from 61.7 to 80.6 across time
segments and from 68.3 to 86.6 across spatial controls. Balanced random
partitions raise the median from 83.4 at \(P_g=2\) to 159.1 at \(P_g=16\), and
the shape partition has \(\widehat{\mathcal C}_R=0.70\) relative to its matched
random control. The topology of the balance is stable, but its numerical scale
is grouping- and window-dependent.

Equation~\eqref{eq:field-preserving-relabelling} shows how attribution can move
while a represented snapshot remains fixed. The staged evidence does not
contain the magnitude grid and ensemble gate needed to reproduce the archived
\(1\%\) sweep, so no binary budgeted verdict or 1404/509/37 trial count is used
here (Appendix~\ref{app:relabel-protocol}). The executable result is nevertheless
decisive within its scope: in this frozen Gaussian representation, the named
groups form a coupled covariance balance concentrated in shared velocity
directions. They are useful geometric descriptors, but not independent,
additive carriers of Reynolds stress.

\subsection{Stability across control representations}
\label{subsec:external-audit}

The strong Gaussian cancellation is not a generic consequence of decomposing a
turbulent field. Two structurally different controls instead show that
component stability is set by the representation geometry, the grouping, and
the permitted coefficient changes. Because the controls act on different
objects, they provide separate scope tests rather than a common binary verdict.

A coherent--incoherent wavelet split provides a realized-field control
\cite{farge1992wavelets,farge2001cvs,schneider2010wavelet}. Membership is
recomputed for every snapshot, and principal angles are evaluated between the
ensemble spans of the resulting coherent and incoherent velocity fields. For
the level-3 \texttt{coif2}, \texttt{db4}, and \texttt{sym4} constructions, the
largest principal cosines are \(0.716\), \(0.742\), and \(0.663\), while
\(\mathcal C_R=1.000\) in all three cases. The largest componentwise change in
the stress half-shares is \(0.0652\). The coherent--incoherent allocation
therefore varies only weakly across the tested wavelet families at the ensemble
level. It
remains a snapshot-relative comparison of realized fields, not a reallocation
within the frozen Gaussian dictionary; reconstruction details and error bounds
are given in Appendix~\ref{app:relabel-protocol}.

The Pope integrity basis provides a complementary coefficient-allocation
control \cite{pope1975moregeneral}. Its ten tensors have effective rank \(8.0\),
and degree grouping gives \(\mathcal C_R=1.000\). With the basis, Gram matrix,
and pooled global coefficient vector fixed, all 2048 seeded perturbations at a
\(1\%\) represented-stress budget retain \(T_8\) as the dominant term. A
targeted least-eigenvalue perturbation at \(5\%\) changes the dominant term to
\(T_2\). The allocation is therefore stable under the declared \(1\%\) random
test, but not independently of perturbation magnitude and direction. This
control concerns global coefficients; it does not test local
invariant-dependent coefficient fields or a trained closure network
\cite{ling2016reynolds}.

Together, the controls show that component allocation need not be uniformly
unstable. They locate the Gaussian result more precisely: cross-group
nonseparability arises from the shared spans and covariance allocation of that
representation, not from decomposing Reynolds stress itself.

\subsection{Stress-divergence bandwidth}
\label{subsec:jhtdb-force}
\label{sec:spectral-enrichment}

Stress fidelity leaves derivative accuracy unconstrained because
differentiation weights the high-wavenumber content of the represented stress:
\(\widehat q_\alpha=-ik_\beta\widehat R_{\alpha\beta}\). On T202,
\(e_R=0.150\) but \(e_q=0.727\), with \(83.9\%\) of the squared divergence
residual at high wavenumber. Five discrete-operator variants give
\(e_q=0.7253\)--\(0.7281\) and high-band fractions \(0.839\)--\(0.840\), ruling
out the derivative implementation as the source.

The error magnitude remains window-dependent. The T202 record has effective
sample size \(12.5\) for \(e_q\), and nested-prefix extrapolation gives
\(E_\infty=0.688\) (Appendix~\ref{app:reproducibility},
Table~\ref{tab:jhtdb-force-scope}). The robust physical result is therefore the
high-wavenumber concentration, not \(e_q=0.727\) as a population value.

For infinite-ensemble HIT, \(R\) is spatially uniform and
\(\nabla\cdot R=0\). Here \(q=-\nabla\cdot R\) is therefore a
derivative-sensitive scope diagnostic, not a nonzero mean Reynolds force.
The residual controls below test whether adding solenoidal fine-scale content
can recover that derivative bandwidth without altering the coarse Gaussian
attribution.

\begingroup
\begingroup
Two residual protocols test the bandwidth diagnosis without changing the
coarse Gaussian attribution. The parameter-selection comparison ranks
residual families against a common represented-variable reference; the
full-field comparison then evaluates the selected family against DNS. Their
errors answer different questions and are not compared across protocols.
\endgroup

\paragraph*{Residual-family comparison.}
Residual-family choice changes stress fidelity but not stress-divergence
fidelity. With the Gaussian representation frozen, each family fits the same
high-pass vorticity residual
\(\epsilon^H=P_{|\mathbf k|\geq13}(\omega'-\widehat\omega_G)\).
Matched candidates are Fourier, Gabor, Gaussian radial-basis functions,
level-2 wavelets, and level-2 wavelet packets, with a Fourier--Gabor union as
composite control.
At \(D_{\rm res}=2048\), Fourier gives the lowest single-family represented
stress error, \(e_R^{\rm repr}=0.122\), against \(0.128\) (Gabor) and
\(0.153\)--\(0.168\) (others)
[Fig.~\ref{fig:residual-family-selection}(a)]. The paired Gabor-minus-Fourier
difference is \(0.00635\) (5th--95th percentile
\([0.00328,0.00747]\), shared origin-block resampling). Fourier is also
exactly graph-orthonormal \((\kappa=1)\).

This ranking does not transfer to stress-divergence fidelity: at matched
budget, all families give \(H^{-1}\) errors between
\(0.318\) and \(0.328\), around the coarse-only value \(0.324\)
[Fig.~\ref{fig:residual-family-selection}(b)]. Fourier is therefore retained
for stress fidelity and conditioning. Within the tested budget, the derivative
error is insensitive to the residual family.

\begin{figure}
  \centering
  \includegraphics[width=\linewidth]
  {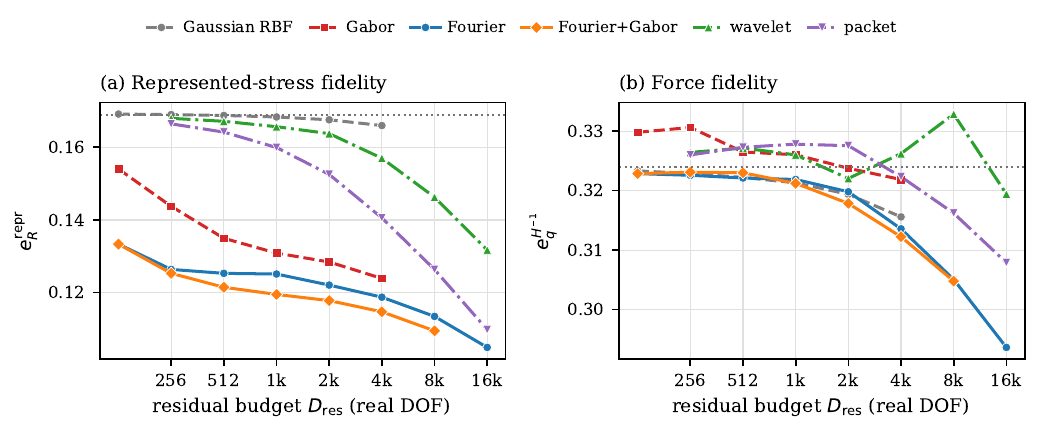}
  \caption{
  Residual-family control on the selection ensemble. (a) Represented-stress
  error against the common represented-variable reference. (b)
  Stress-divergence error in the \(H^{-1}\) norm. Lower is better.}
  \label{fig:residual-family-selection}
\end{figure}

\paragraph*{Full-field bandwidth enrichment.}
Adding high-pass Fourier content improves stress more readily than stress
divergence. The selected residual augments the full represented velocity as
\[
  \widehat{\mathbf u}_s
  =
  \overline{\mathbf u}_{s,h}
  +\Phi_GA_{s,:}^{\mathsf T}
  +F\mathbf b_s,
\]
where \(\Phi_G=\Phi\) is the frozen Gaussian velocity dictionary and
\(F\mathbf b_s\) the fitted residual restricted to
\(|\mathbf k|\geq13\); the Gaussian-envelope attribution is unchanged. The
full-field protocol uses \(R_{\rm ref}=\operatorname{Cov}(u')\) rather than
the represented-variable reference above. We report \(e_R\), \(e_q\), and
\[
  e_q^{H^{-1}}
  =
  \frac{\|q^{\rm hyb}-q^{\rm ref}\|_{H^{-1}}}
       {\|q^{\rm ref}\|_{H^{-1}}},
  \qquad
  \|f\|_{H^{-1}}^2
  =
  \sum_{\mathbf k\ne0}|\mathbf k|^{-2}|\widehat f(\mathbf k)|^2 .
\]

\begin{table}
  \caption{Fourier residual enrichment.}
  \label{tab:hybrid-force-closure}
  \centering
  \footnotesize
  \setlength{\tabcolsep}{3.5pt}
  \begin{ruledtabular}
  \begin{tabular}{lcccc}
    Representation\footnote{Results use the selection ensemble. HP and FB
    denote high-pass and full-band Fourier, respectively.}
    & \(D_F\)\footnote{Number of residual coefficients.}
    & \(e_R\) & \(e_q\) & \(e_q^{H^{-1}}\) \\
    \colrule
    Gaussian coarse & 0 & 0.112 & 0.867 & 0.292 \\
    \(\,+\) HP Fourier (retained) & 2048 & 0.084 & 0.816 & 0.281 \\
    \(\,+\) HP Fourier & 16384 & 0.073 & 0.771 & 0.256 \\
    FB Fourier & 28672 & 0.058 & 0.714 & 0.179 \\
    HP Fourier only & 4096 & 0.602 & 0.924 & 0.992 \\
  \end{tabular}
  \end{ruledtabular}
\end{table}

\begingroup
At \(D_F=2048\), enrichment reduces \(e_R\) by \(25\%\) but raw and
\(H^{-1}\) stress-divergence errors by only \(5.9\%\) and \(3.8\%\).
Increasing the high-pass budget eightfold raises those reductions only to
\(11.1\%\) and \(12.3\%\); the high-pass-only control fails because it removes
the coarse field. Full-band Fourier reaches \(e_q^{H^{-1}}=0.179\), but then
competes at every scale rather than supplying a distinct residual channel.
Residual-family substitution is therefore not the bottleneck; scale coverage
is. Improving derivative fidelity requires broader bandwidth without
discarding the local solenoidal representation or its complete covariance
accounting.
\endgroup

\endgroup

\section{Reynolds-stress structure in channel flow}
\label{sec:channel-extension}

\begingroup
Wall-normal inhomogeneity makes stress location dynamically consequential:
Reynolds shear contributes to production only where it overlaps the mean
velocity gradient. The channel case therefore asks three connected questions.
Does the representation satisfy the wall constraints, where does it place the
represented stress, and how is that stress assembled among its components?
The answers distinguish exact kinematics from stress fidelity and, in turn,
stress fidelity from component attribution. Wall and outer regions are decoded
independently, with neither covering \(y^+\simeq284\)--\(302\); all conclusions
below are therefore region-local.
\endgroup

\subsection{Boundary compatibility and stress fidelity}
\label{subsec:channel-coordinate}
\label{subsec:channel-fidelity}

Exact wall constraints define an admissible velocity field, but do not by
themselves determine its Reynolds stress. We represent the fluctuation
\(\mathbf u'_s=\mathbf u_s-\mathbf U(y)\) with curl-lifted Gaussian envelopes.
For envelope \(g_i\), Cartesian direction \(\mathbf e_c\), and wall distance
\(d_w\),
\begin{equation}
  \mathbf A_{i,c}=w(d_w)g_i\mathbf e_c,
  \qquad
  \boldsymbol\phi^{\,w}_{i,c}=\nabla\times\mathbf A_{i,c},
  \qquad
  \widehat{\mathbf u}'_s
  =\sum_{i,c}a_{s,i,c}\boldsymbol\phi^{\,w}_{i,c},
  \label{eq:channel-wall-representation}
\end{equation}
where \(w(d_w)=1-\exp[-(d_w/\ell_w)^2]\) and
\(\ell_w=15\delta_\nu\). The curl enforces solenoidality, while
\(w(0)=w'(0)=0\) gives no-slip for every coefficient vector: tangential and
wall-normal velocity components scale as \(O(d_w)\) and \(O(d_w^2)\),
respectively.

For the channel flow from the Johns Hopkins Turbulence Databases (JHTDB)
\cite{jhtdb,graham2016web,lee2015direct} at
\(\mathrm{Re}_{\tau}\approx1000\), separate frozen dictionaries represent the
wall and outer regions. The wall dictionary uses the taper above and the outer
dictionary sets \(w=1\). Their retained \(23^3\)-envelope (\(K23\)) geometries
are evaluated at eight previously unused times by refitting amplitudes only;
the complete sampling and selection protocol is given in
Appendix~\ref{app:reproducibility}.

The resulting stress is substantially more accurate in the outer region than
near the wall. For
\(R_{ij}(y)=\langle u_i'u_j'\rangle_{s,x,z\mid y}\), frozen \(K23\) gives
\((e_R,e_{uv})=(0.0661,0.0269)\) outside and \((0.129,0.123)\) near the wall;
both dictionaries satisfy the prescribed conditioning gate. The retained
field also preserves the principal stress directions and local componentality,
with its largest discrepancies again near the wall
(Appendix~\ref{app:channel-tensor-state}). Thus the main loss is concentrated
where the turbulent scales shorten and wall-normal gradients sharpen.

The controls separate this resolution loss from conditioning alone. A denser
\(K26\) lattice lowers the parameter-selection error but violates the
conditioning gate, whereas a matched uniform-knot B-spline representation is
better conditioned but less accurate, with \(e_R=0.401\)--\(0.440\)
(Appendix~\ref{app:reproducibility}, Table~\ref{tab:channel-fidelity}). These
comparisons support the retained adaptive envelopes within the tested local
construction; they do not rank them against a wall-resolving
Fourier--Chebyshev or wall-clustered-spline representation.

With the admissible field and its stress fidelity established, the
next question is whether the represented shear is placed where it can extract
energy from the mean flow.

\subsection{Wall-normal placement of Reynolds shear}
\label{subsec:channel-production-interface}

\begin{figure*}
  \centering
  \includegraphics[width=\textwidth]
  {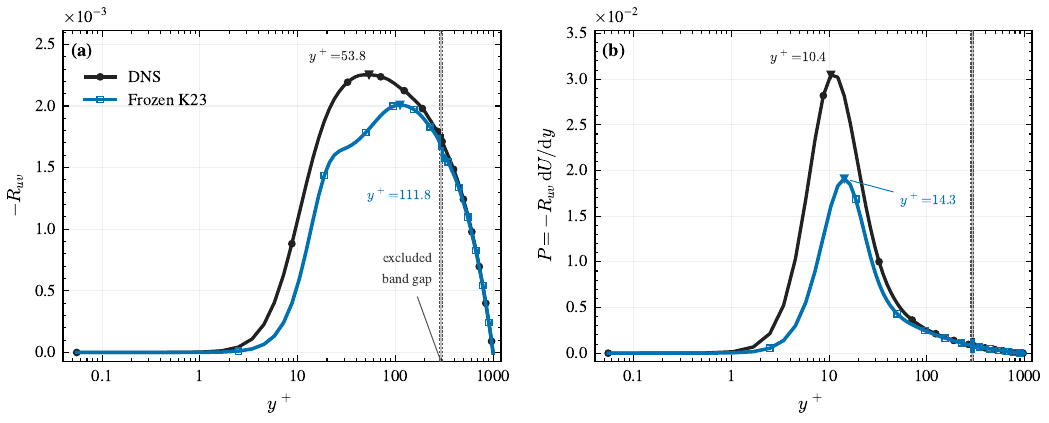}
  \caption{Wall-normal shear and production for the frozen \(K23\)
  representation. (a) DNS and represented \(-R_{uv}\). (b)
  \(P=-R_{uv}\,\mathrm dU/\mathrm dy\), evaluated with the DNS mean gradient.
  Dotted lines delimit the gap between the two regional representations.}
  \label{fig:channel-shear-production}
\end{figure*}

Near-wall production is limited more by the position of Reynolds shear than by
its peak magnitude. Frozen \(K23\) retains \(89.1\%\) of the DNS peak of
\(-R_{uv}\), but moves it from \(y^+=53.8\) to \(111.8\)
[Fig.~\ref{fig:channel-shear-production}(a)]. The displaced stress therefore
samples a weaker mean gradient. With the DNS mean profile held fixed, the
production \(P=-R_{uv}\,\mathrm dU/\mathrm dy\) peaks at \(y^+=14.3\), rather
than \(10.4\), and has wall-region error \(e_P=0.377\) and integrated ratio
\(0.761\). The corresponding outer-region values are \(0.0313\) and \(0.975\)
[Fig.~\ref{fig:channel-shear-production}(b)].

This separation between stress magnitude and stress location is the main
channel-flow result. A field may satisfy no-slip and retain the leading stress
amplitude while placing momentum flux too far from the wall to reproduce its
production. Trimming four face-adjacent points changes \(e_R\) by at most
\(1.27\%\), so the discrepancy does not originate in the edge cells. The two
regional decoders nevertheless have order-one shared-face jumps
(Appendix~\ref{app:reproducibility}). Improved near-wall fidelity therefore
requires both finer wall-normal placement and coupling across regional
interfaces.

Stress placement controls production; the remaining attribution
question concerns how the represented shear is assembled among the frozen
groups.

\subsection{Self--cross covariance of Reynolds shear}
\label{subsec:channel-attribution}

The represented Reynolds shear is not an additive sum of independently
stress-carrying groups. It is a small residual between much larger positive
self contributions and negative cross contributions, extending the relational
stress organization found in HIT to wall-normal inhomogeneity. We partition
the frozen \(K23\) wall envelopes independently by mean-width and
centre-wall-distance terciles. In each case,
Eq.~\eqref{eq:closed-group-accounting} closes all
\(R_{uv}^{(p,q)}(y)=\langle u'_p v'_q\rangle_{s,x,z\mid y}\) blocks exactly to
the represented shear.

Normalize the integrated self and cross terms by the net shear and define
\(\zeta_{uv}=|\eta_{uv}^{\rm cross}|/\eta_{uv}^{\rm self}\). The width and
wall-distance partitions give
\[
  \begin{aligned}
  (\eta_{uv}^{\rm self},\eta_{uv}^{\rm cross},\zeta_{uv})_{\rm width}
  &= (12.208,-11.208,0.918),\\
  (\eta_{uv}^{\rm self},\eta_{uv}^{\rm cross},\zeta_{uv})_{\rm wall}
  &= (9.787,-8.787,0.898).
  \end{aligned}
\]
Cross covariance therefore removes \(89.8\%\)--\(91.8\%\) of positive self
contributions that are \(9.8\)--\(12.2\) times the net shear.

This signed balance survives every prespecified temporal and spatial deletion
and all \(2000\) nested resamples at time-block lengths \(1\), \(2\), and \(4\).
The cancellation fraction remains \(0.909\)--\(0.923\) for the width partition
and \(0.880\)--\(0.917\) for the wall-distance partition; alternative nonempty
partitions retain \(\zeta_{uv}\geq0.861\). Wall-normal inhomogeneity therefore
does not remove the need for complete cross covariance: in this region-local
representation, Reynolds shear is carried by relations among groups rather
than by their isolated self terms. The tercile labels remain coordinates of
the frozen dictionary, not uniquely identified physical wall structures.

\section{Discussion}
\label{sec:discussion}

\begingroup
The central physical result is that a localized description of vortical motion
need not yield an additive inventory of stress-carrying structures. Freezing
the decoder makes coefficients and covariance blocks comparable across
snapshots, but it cannot separate groups whose decoded velocity spans overlap.
For such groups, the complete self--cross balance is the well-defined
contribution to Reynolds stress; an isolated group share need not be. This
relational interpretation preserves the physical value of spatially
addressable morphology without turning its labels into unique flow objects.

The evidence identifies the mechanism of that nonadditivity. Near-shared
sheet--tube directions contain almost all of the measured cancellation.
Orthogonal regrouping removes span redundancy, while amplitude row-shuffling
removes inter-atom correlation; both return the cancellation index to unity.
A second nonorthogonal coordinate develops stronger opposing blocks as its
overlap increases at fixed field and stress error. Geometry therefore supplies
interchangeable velocity directions, and coefficient covariance determines how
strongly those directions are jointly populated. Neither morphology nor
reconstruction error alone explains the stress balance. Related signed
balances in scale-decomposed stress transport suggest that this distinction may
also matter beyond the present grouping
\cite{kawata2018inverse,kawata2019scaleinteractions}.

This result also explains why the localized representation is retained when
orthogonal coordinates are available. Fourier and POD coordinates separate
their chosen global modes, but they do not provide the same spatially
addressable anisotropic support. The Gaussian representation retains that
local description at the cost of overlap. The appropriate comparison is
therefore not a universal ranking of bases, but a tradeoff among fidelity,
locality, conditioning, and component separability. Any physical attribution
made in a local nonorthogonal coordinate must report all four.

The fidelity results expose two further constraints on turbulent transport.
Exact covariance propagation, solenoidality, and no-slip determine admissible
fields, but not derivative bandwidth or wall-normal stress placement. In HIT,
differentiation reveals unresolved high-wavenumber stress content. In the
channel, a displaced Reynolds-shear peak weakens production even though most of
its peak magnitude is retained. A more complete local representation must
therefore resolve not only stress magnitude, but also its gradients, its
near-wall position, and continuity across regional interfaces.

The control representations sharpen rather than universalize these findings.
The coherent--incoherent wavelet result concerns snapshot-dependent realized
fields; the integrity-basis result concerns one pooled global coefficient
vector. Neither is a trained local closure or a relabelling within the Gaussian
span. The unavailable Gaussian perturbation ledger also prevents a binary
\(1\%\) stability verdict. More generally, stability under a declared
perturbation is necessary, but not sufficient, for causal, unique, or
representation-independent meaning
\cite{brunton2020machine,taira2020modal,ling2016reynolds,
locatello2019challenging,karniadakis2021physics}.

The numerical evidence covers one bandwidth-limited JHTDB HIT regime and one
channel case at \(\mathrm{Re}_\tau\approx1000\). T202 is a correlated
single-origin record with effective sample size near ten, and the channel
representation contains independently decoded regions separated by an
unrepresented band. Within these bounds, the transferable conclusion is
strong: stress fidelity, complete covariance accounting, spatial placement,
and component separability are distinct properties. Keeping them distinct
turns a local nonorthogonal representation into a quantitative description of
coupled turbulent transport rather than an unsupported census of structures.
\endgroup

\section{Conclusions}
\label{sec:conclusion}

\begingroup
Accurate reconstruction does not determine how a localized nonorthogonal flow
representation assigns Reynolds stress to its components. By combining
decoder geometry with complete self--cross covariance accounting, this study
separates the fidelity of the represented stress from the separability of its
component fields.

In the frozen solenoidal Gaussian representation, more than half of the sheet
span is near-shared with the tube span, and products within the shared
directions carry nearly all of the covariance cancellation. Removing
cross-group covariance increases the DNS-referenced stress error by more than a
factor of fifty. The supported physical conclusion is cross-group
nonseparability within this representation: localized vortical components
contribute to Reynolds stress through coupled fluctuations, not as independent
owners of additive stress shares. The channel case extends the same relational
organization to wall-normal inhomogeneity and shows that retaining stress
magnitude cannot compensate for misplaced Reynolds shear in turbulent
production.

Exact kinematic constraints and exact covariance accounting remain distinct
from derivative resolution, wall-normal placement, and component
separability. The present conclusion rests on executable span, covariance, and
control evidence; it assigns no binary verdict to the unrecoverable Gaussian
perturbation sweep. Treating representation geometry, grouping, ensemble, and
perturbation family as part of each attribution claim provides a rigorous path
for using localized representations to expose how turbulent momentum flux is
assembled.
\endgroup

\begingroup
\section*{Supplementary Material}

The Supplemental Material contains the nomenclature (Sec.~S1), the matched
represented-variable control (Sec.~S2), and file-level evidence provenance
(Sec.~S3).
\endgroup

\begingroup
\section*{Author Declarations}

\subsection*{Conflict of Interest}

The authors have no conflicts to disclose.

\subsection*{Author Contributions}

\textbf{Peng Cheng}: Conceptualization (lead); Methodology (lead); Software
(lead); Formal analysis (lead); Investigation (lead); Writing -- original draft
(lead).
\textbf{Ming Jian}: Methodology (supporting); Validation (equal); Writing --
review \& editing (equal).
\textbf{Wen Tong}: Supervision (lead); Resources (lead); Writing -- review \&
editing (equal).
\endgroup

\section*{Data Availability Statement}

The DNS data used in this study are publicly available through the Johns
Hopkins Turbulence Databases (JHTDB), including the forced isotropic turbulence
and turbulent channel-flow datasets. The raw DNS fields are not redistributed.
The analysis code and supporting evidence are available from the corresponding
author upon reasonable request.

\clearpage
\appendix
\section{Controls on representation design}
\label{app:design-ablations}

Section~S2 of the Supplemental Material establishes the choice of vorticity as
the represented variable. The matched control below then fixes that variable
and isolates how directional envelope support changes fidelity and numerical
independence.

\subsection{Directional support and conditioning}
\label{app:geometry-family-evidence}

\begingroup
Direction-dependent widths improve both stress fidelity and conditioning;
free rotation gives no separately resolved gain at this budget.  The matched
comparison fixes
\((N_{\mathrm{geom}},N_{\mathrm{opt}})=(64,40)\) and determines the
learned-isotropic (G1), axis-aligned (G2), regularized oriented (G3), and
penalty-free oriented (G4) families before reading the reserved and T202
ensembles. Table~\ref{tab:fair-geometry-ablation} reports the matched
geometry-family comparison. Paired origin-block resampling supports the
reserved-data comparison; T202 remains descriptive.
\endgroup

\begin{table}
\caption{Geometry-family control.}
\label{tab:fair-geometry-ablation}
\footnotesize
\setlength{\tabcolsep}{3pt}
\begin{ruledtabular}
\begin{tabular}{lllccccc}
Family\footnote{All families use deterministic \(16^3\) centers,
\(N_G=4096\), \(M_b=12288\), and the common initialization, objective,
optimizer, and 40-update budget of
Appendix~\ref{app:geometry-family-evidence}.}
& geometric freedom
& penalty\footnote{\(\alpha_\star=729.699\) is the common retained coherence
weight for G1--G3; G4 sets \(\alpha=0\).}
& selection $e_R$\footnote{Errors are mean \(\pm\) sample standard deviation
over five paired geometry-ensemble resamples.}
& reserved $e_R$ & T202 $e_R$
& $\kappa(K_G)$\footnote{Gram statistics are means over the five selected
geometry states.}
& $r_{\rm eff}(K_G)$ \\
\colrule
G1 & isotropic width & $\alpha_\star$
& $0.1469\pm0.0001$ & $0.1480\pm0.0003$ & $0.1995\pm0.0003$
& $6.52\times10^7$ & $2601$ \\
G2 & axis-aligned widths & $\alpha_\star$
& $0.1380\pm0.0003$ & $0.1395\pm0.0003$ & $0.1863\pm0.0005$
& $2.23\times10^6$ & $2757$ \\
G3 & oriented widths & $\alpha_\star$
& $0.1389\pm0.0003$ & $0.1403\pm0.0003$ & $0.1877\pm0.0004$
& $1.49\times10^6$ & $2709$ \\
G4 & oriented widths & $0$
& $0.1334\pm0.0002$ & $0.1354\pm0.0001$ & $0.1792\pm0.0004$
& $3.28\times10^7$ & $1439$ \\
\end{tabular}
\end{ruledtabular}
\end{table}

Relative to G1, G2 lowers reserved-ensemble \(e_R\) from \(0.1480\) to \(0.1395\) and T202
\(e_R\) from \(0.1995\) to \(0.1863\); all five paired comparisons agree in
sign.  G2 also reduces \(\kappa(K_G)\) by a factor of \(29\) and raises
effective rank from \(2601\) to \(2757\).  G3 differs from G2 by less than
\(\epsilon_R=0.005\), leaving free orientation separately unresolved under the
retained penalty.  Removing that penalty lowers \(e_R\) but makes the Gram
\(22\) times more ill-conditioned and reduces effective rank by \(47\%\).

Directional support is therefore resolved as a useful degree of freedom,
whereas unrestricted overlap trades a small fidelity gain for a substantial
loss of numerical independence. The retained oriented state follows the
separate fixed-object stability criterion of
Sec.~\ref{subsec:verify-geometry-selection}.

\section{Numerical and statistical protocols}
\label{app:reproducibility}

\begingroup
This appendix fixes the numerical objects, data roles, statistical treatment,
and perturbation evidence behind the main conclusions. It distinguishes fully
specified procedures from aggregate-only evidence so that no physical claim is
stronger than the available record.
\endgroup

\subsection{Numerical construction}

\paragraph*{Discrete operators and normalization.}
Periodic displacements are wrapped to the nearest image. A sampled, freely
oriented envelope need not be smooth across its periodic cut planes, so the
compatibility statements in Sec.~\ref{subsec:vorticity-decoding} are discrete
spectral identities. The target curl and Gaussian curl lift share the
Nyquist-zeroed derivative wavevector \(\mathbf k_d\); the inverse
Laplacian retains the full \(|\mathbf k|^2\).

Before lifting, each envelope is mean-centered and root-mean-square
normalized. We use
\[
  \mathbf g_{i,c}^{\theta}
  =
  \frac{\nabla_h\times
    (\check g_i^\theta\mathbf e_c)}
  {\|\nabla_h\times
    (\check g_i^\theta\mathbf e_c)\|_W},
  \qquad c=1,2,3 ,
\]
the operator \(\mathcal N_W\) of
Eq.~\eqref{eq:curl-lifted-atoms}. Mean subtraction and scalar rescaling
preserve the curl direction; the final normalization gives every atom unit
design norm, so \(N_G\) envelopes generate \(M_b=3N_G\) normalized columns.
Reported amplitudes multiply these columns; the raw-column coefficient is
\(a_{s,i,c}/\nu_{i,c}\), with \(\nu_{i,c}\) the pre-normalization design norm.

\paragraph*{Geometry determination.}
The \(N_G=4096\) centers are fixed on a deterministic \(16^3\) lattice.
The retained representation selected in
Sec.~\ref{subsec:verify-geometry-selection} uses
\((N_{\mathrm{geom}},N_{\mathrm{opt}})=(64,40)\). Its geometry remains fixed,
and new snapshots refit amplitudes only.
Independent geometry-family comparisons start from
\(\sigma_{i,1}=\sigma_{i,2}=\sigma_{i,3}=5\Delta x\) and \(Q_i=I\).

Widths are optimized in log coordinates, and oriented families add
three-component Rodrigues increments with re-orthonormalization.
Adam uses width and rotation step sizes \(0.03\) and \(0.015\),
\((\beta_1,\beta_2)=(0.9,0.999)\), \(\epsilon=10^{-8}\), and unit RMS
gradient clipping. Isotropic, axis-aligned, and oriented envelopes give
\(N_G\), \(3N_G\), and \(6N_G\) optimized degrees of freedom, and every family
fits the same \(M_b=12288\) amplitudes per snapshot. The independent comparison
records updates \(10,20,30,40\); the retained trajectory records
\(12,40,80,160\), retaining update \(40\).

\paragraph*{Metrics, constraints, and amplitudes.}
The graph metric uses \(\ell=0.25\Delta x\), \(N_s=32768\)
deterministically sampled points (seed \(17\)), and relative ridge
\[
  \lambda_{\mathrm{ridge}}
  =
  10^{-8}\operatorname{mean}
  [\operatorname{diag}(K_G)] .
\]
The geometric term in Eq.~\eqref{eq:learning-objective} is enforced by
projection: widths lie in \([\Delta x,16\Delta x]\), aspect ratio is at most
\(8\), and centers do not move. The Gram threshold is
\(\tau_{\mathrm{Gram}}=0.5\), and its
retained calibration assigns initial objective share \(0.1\); the
independent comparison fixes \(\alpha=729.6990856\) for G1--G3 and
\(\alpha=0\) only for G4. The stress-aware coefficient is
\(\beta_{\mathrm{stress}}=0\) throughout.
A separate float64 check compares Eq.~\eqref{eq:covariance-lift} with the
direct covariance of the decoded velocity fields, with relative discrepancy
\(1.5\times10^{-16}\).

\paragraph*{Computational cost.}
The float64 sampled design has size \((3N_s)\times M_b\); its Cholesky
factorization is reused at fixed geometry. Table~\ref{tab:matched-footprint}
shows that the Gaussian construction is substantially more expensive than the
matched Fourier representation; it is retained for spatial addressability, not
computational economy.

\begin{table}
\caption{\label{tab:matched-footprint}
Matched computational cost.}
\footnotesize
\setlength{\tabcolsep}{4pt}
\begin{ruledtabular}
\begin{tabular}{lcccc}
& \multicolumn{2}{c}{Median time (s)}
& \multicolumn{2}{c}{Peak GPU (GB)} \\
\cline{2-3}\cline{4-5}
Stage\footnote{Both representations use the same machine, GPU, float64
precision, and \(S=128\) snapshots. Times are stage medians; memory is peak
allocation.} & Gaussian & Fourier & Gaussian & Fourier \\
\colrule

\multicolumn{5}{l}{\emph{One-time setup}} \\
1\ \ Dictionary construction
& \(441.2\) & \(0.4343\) & \(0.008\) & \(4.93\) \\
2\ \ Sampled design and Gram\footnote{
Fourier requires no sampled Gram or factorization because its graph-weighted
normal equations are diagonal.}
& \(319.6\) & N/A & \(0.008\) & --- \\
3\ \ Factorization\footnotemark[1]
& \(0.0230\) & N/A & \(5.63\) & --- \\
\colrule

\multicolumn{5}{l}{\emph{Per snapshot}} \\
4\ \ Single-snapshot solve
& \(0.0252\) & \(0.000250\) & \(13.54\) & \(6.43\) \\
5\ \ Velocity decode
& \(0.00336\) & \(0.00132\) & \(20.49\) & \(6.49\) \\
\colrule

\multicolumn{5}{l}{\emph{Stress assembly, common procedure}\footnote{
The two arms use the same numerical operations and agree within repeat spread.}} \\
6\ \ \(S=1\)
& \(0.000164\) & \(0.000163\) & \(6.48\) & \(6.49\) \\
6\ \ \(S=128\)
& \(0.00279\) & \(0.00280\) & \(7.48\) & \(7.48\) \\
\colrule

\multicolumn{5}{l}{\emph{Totals}} \\
Setup (stages 1--3)
& \(760.8\) & \(0.4343\) & & \\
Marginal per snapshot (4--5)
& \(0.0286\) & \(0.00157\) & & \\
Cost ratio\footnote{
Common-pipeline total is setup plus \(S\) marginal solves and shared stress
assembly. The larger Gaussian marginal cost precludes a positive break-even.}
at \(S=1\) & \multicolumn{2}{c}{\(1745\times\)} & & \\
Cost ratio\footnotemark[3] at \(S=128\)
& \multicolumn{2}{c}{\(1198\times\)} & & \\
\colrule

\multicolumn{5}{l}{\emph{Gaussian-only capability, excluded from every
total}\footnote{
Fourier has no matched local-attribution stages.}} \\
7\ \ Family assignment
& \(0.00115\) & --- & \(6.46\) & --- \\
8\ \ Pair covariance attribution
& \(0.0376\) & --- & \(10.28\) & --- \\
9\ \ Bootstrap attribution
& \(7.817\) & --- & \(13.94\) & --- \\

\end{tabular}
\end{ruledtabular}
\end{table}

\subsection{HIT sampling and statistical treatment}
\label{subsec:jhtdb-protocol}

\paragraph*{HIT ensembles.}
The HIT analysis uses quarter-box cutouts from a JHTDB ensemble
\cite{jhtdb}. The source is \texttt{isotropic1024coarse}; each \(256^3\) cutout is
sampled at stride four onto a \(64^3\) grid. Because the cutout is not
periodic, the same periodic Leray projection is applied to the reference and
represented fields before spectral fitting and Biot--Savart decoding. This is
a common window treatment, not a surrogate for the DNS dynamics.

The database reports \(\mathrm{Re}_\lambda\approx433\) \cite{jhtdb}. With
\(\nu=1.85\times10^{-4}\) and \(\langle\epsilon\rangle=0.0928\), the
Kolmogorov length is
\(\eta_K=(\nu^3/\langle\epsilon\rangle)^{1/4}=2.87\times10^{-3}\).
Because \(\eta_K/\Delta x=0.117\), the analysis spacing is
\(\Delta x\approx8.5\,\eta_K\). The sampled fields are therefore strongly
bandwidth limited relative to the DNS, and every HIT velocity and stress
comparison refers to this common resolved bandwidth.

Per-snapshot preprocessing consists of deterministic stride-four sampling
where required, spectral Leray projection, spectral curl, and float64 storage.
The canonical times are
\[
  \{0,667,1333,2000,2667,3333,4000,4667\}.
\]
Geometry determination uses \(31\) origins by \(8\) times (\(S=248\));
parameter selection and selection-free reserved evaluation each use
\(16\times8\) (\(S=128\)). Their origins are mutually disjoint and disjoint
from geometry determination, although both reuse its eight time indices; hence
the roles are snapshot-disjoint but not time-index-disjoint. T202 uses 202
consecutive snapshots at one separate origin. Of its time indices, 199 are new
and three are shared with the canonical set, while every origin--time pair is
new. The main 32-origin transfer result combines selection and reserved data and
is therefore a secondary pooled aggregate, not a selection-free estimate. The
\(S=504\) union contains the 31 geometry, 16 selection, and 16 reserved origins
and is a sampling-consistency ensemble, not a transfer estimate. A fail-closed
split verification reproduces these roles with no origin or snapshot overlap
beyond the shared time indices stated above and confirms common preprocessing
across roles (Sec.~S3). Every ensemble is centered separately; pooled
covariances and \(R^{\mathrm{DNS}}\) are therefore re-estimated rather than
assembled from subset values.

Each of the five geometry-comparison replicates uses \(8\) origins by \(8\)
times from the 31-origin pool. Seed \(0\) uses the retained nested prefix;
seeds \(1\)--\(4\) use deterministic origin permutations generated by
\(1001,\ldots,1004\).

\paragraph*{T202 temporal dependence.}
The T202 protocol uses \(2000\) paired circular-block bootstrap replicates
(seed \(20260722\)), each recomputing centering, reference and
represented covariances, stress divergence, and all six family-pair blocks.
Block lengths \(L\in\{13,26,44,51\}\) are tested; the rule
\(L^\star=\lceil2\tau_{\mathrm{corr,max}}\rceil\) selects \(L^\star=44\).
Effective sample sizes are \(9.3\) for \(e_R\), \(12.5\) for
\(e_q\), and approximately \(7\)--\(10\) for individual volume-mean stress
components. These values fail the prespecified interval-reporting gates, so
the T202 headline errors remain deterministic diagnostics conditional on that
origin and finite window.

\paragraph*{Finite-window and bandwidth sensitivity.}
Table~\ref{tab:jhtdb-force-scope} records the T202 bandwidth diagnostic and its
window, measure, reweighting, and fixed-residual controls.

\begin{table}
\caption{Stress-divergence controls.}
\label{tab:jhtdb-force-scope}
\centering
\footnotesize
\begin{ruledtabular}
\begin{tabular}{lc}
Diagnostic\footnote{T202 uses the retained
\((N_{\rm geom},N_{\rm opt})=(64,40)\) representation on one unseen origin
and 202 times. Here
\(e_q=\|q^{\mathrm{dep}}-q^{\mathrm{DNS}}\|_2/
\|q^{\mathrm{DNS}}\|_2\), and \(f_q^B\) is the fraction of squared
stress-divergence residual in band \(B\). The seam-free control uses 128
periodic whole-box snapshots; the sample-floor fit uses nested T202 prefixes.}
& Value \\
\colrule
T202 full-grid stress error \(e_R\) & \(0.150\) \\
T202 full-grid stress-divergence error \(e_q\) & \(0.727\) \\
Squared stress-divergence residual shares
\((f_q^{\mathrm L},f_q^{\mathrm M},f_q^{\mathrm H})\)
& \((0.017,\,0.144,\,0.839)\) \\
Trim-4 interior \((e_R,e_q)\) & \((0.095,\,0.833)\) \\
\colrule
Seam-free whole-box \(e_q\) & \(0.815\) \\
Nested-prefix extrapolated floor \(E_\infty\) & \(0.688\) \\
Oracle reweighting: \(e_q\) reduction & \(0.18\%\) \\
Best fixed residual representation: \(e_q\) reduction & \(0.31\%\) \\
\end{tabular}
\end{ruledtabular}
\end{table}

Because \(\mathcal C_R\) divides strongly cancelling blocks by their small net,
the analysis emphasizes block numerators, signs, and dominant-pair retention.
All tested block lengths retain the pair signs and the dominant sheet--tube
cross block, whereas a four-segment comparison moves \(\mathcal C_R\) from
\(61.7\) to \(80.6\). The sign structure is stable; the normalized magnitude
is window-dependent.

\paragraph*{Measure and representation sensitivity.}
The shape ratio \(1.5\), aspect cutoff \(2.5\), scale cutoffs
\(2\Delta x\) and \(4\Delta x\), and high-lift fraction were fixed before
T202.  Interior trims, Tukey weights, shorter windows, and a periodic
whole-box measure preserve positive self blocks, negative cross blocks, and
the dominant opposing pair without refitting; cancellation magnitude remains
measure-dependent.

Threshold sweeps leave the decoded field, total stress, and exact accounting
unchanged. Aspect and scale partitions are stable, whereas shape and lift
labels are threshold-sensitive. At principal cosine \(0.95\), the atom-span
analysis finds \(1703\) near-shared sheet--tube directions, or \(55\%\) of the
sheet rank.

\paragraph*{Selection criteria.}
The prescribed candidate grid is
\[
N_{\mathrm{geom}}\in\{15,64,128\},
\qquad
N_{\mathrm{opt}}\in\{12,40,80,160\}.
\]
The product \(B=N_{\mathrm{geom}}N_{\mathrm{opt}}\) measures geometry-
determination exposure, not runtime. A deterministic replay at
\(N_{\mathrm{geom}}=64\) also evaluates updates \(56\) and \(60\); the full
trajectory is retained for this comparison.
The original joint gate required
\(e_R\le e_{R,\min}+0.005\), zero near-null Gram dimension,
\(\kappa_+\le10^9\), covariance and pair-accounting residuals below
\(10^{-11}\), pair cosines at least \(0.95\) under origin-block resampling and
\(0.90\) across time windows, closed-vector cosine at least \(0.95\), sign
consistency at least \(0.95\), and unchanged sign topology and dominant pair.
No geometry state passes this full rule.

The secondary criterion retaining
\((N_{\mathrm{geom}},N_{\mathrm{opt}})=(64,40)\) requires minimum closed
group share above \(-10^{-3}\),
fifth-percentile bootstrap closed-vector cosine at least \(0.95\), cosine to
the retained closed vector at least \(0.95\), and unchanged dominant pair.
A fidelity gain is material only when \(\Delta e_R\ge0.005\) and its paired
95\% interval has positive lower bound. This replay uses \(500\) shared
origin-block draws with seed \(20260718\).  These gates are prespecified, not
post-hoc.

\subsection{Algebraic sign constraint}
\label{app:covariance-sign}

The sign of each self alignment is constrained before any flow-specific
measurement. With \(W_6=(1,1,1,2,2,2)\), the packed-tensor product used in
Eq.~\eqref{eq:pair-alignment} is \(\operatorname{tr}(AB)\). The coefficient
block \(C_{pp}\) is a principal submatrix of the positive-semidefinite
covariance \(C_A\); hence both \(R^{(p,p)}(\mathbf x)\) and
\(R^{\mathrm{coef}}(\mathbf x)\) are positive semidefinite. Their tensor
contraction is nonnegative pointwise, and therefore
\(\eta_{pp}^{R}\ge0\) after spatial integration. Exact pair accounting gives
\(\sum_{p\le q}\eta_{pq}^{R}=1\), so
\[
  \sum_p\eta_{pp}^{R}>1
  \quad\Longrightarrow\quad
  \sum_{p<q}\eta_{pq}^{R}<0.
\]
This identity fixes the sign of the total cross contribution once the self sum
exceeds unity; it does not determine the magnitude, the signs of individual
cross blocks, or their spatial and geometric organization.

\subsection{Perturbation protocol and evidence boundary}
\label{app:relabel-protocol}

The available atom-span record supplies aggregate Gaussian relabelling counts
and a principal-overlap summary. It records 1404 total field-preserving and
group-preserving trials, 509 trials admitted by its declared \(1\%\) gates, and
37 admitted trials that break an attribution criterion. It also records exact
recovery of the frozen baseline and passing sanity controls at tolerance
\(10^{-10}\). Trial-level perturbations and the complete generation protocol
are not available in the submission evidence. Consequently, these counts are
aggregate evidence rather than an independently reproducible result.

\paragraph*{Recoverable Gaussian protocol.}
The frozen objects are the Gaussian decoder, the blob--sheet--tube grouping,
the kinetic-energy inner product \(M_u\), the group spans, their principal
vectors, and the baseline group fields. The report exposes the complete
principal-cosine set
\[
 \tau_{\mathrm{PA}}\in\{0.75,0.90,0.95,0.99,0.999\}
\]
and the allocation set
\[
 \alpha_{\mathrm{alloc}}\in\{0,0.25,0.50,0.75,1\}.
\]
For each selected pair and parameter tuple, Eq.~\eqref{eq:field-preserving-relabelling}
constructs snapshot-level coefficients \(\delta_{j,s}\). Mode A adds equal and
opposite increments, so every total represented snapshot, its centered
covariance, and \(e_R\) remain fixed. Mode B applies the \(M_u\)-orthogonal
group projections before the update, so group membership is retained and the
total represented field and stress may change. The report defines \(t=0\) as
the baseline attribution and \(t=1\) as complete reallocation, and evaluates
the largest gate-passing \(t\) for each exposed parameter combination.

The magnitude grid \(\mathcal T\subset[0,1]\) between those endpoints is not
listed in the available record, which also does not show the exact
Cartesian-product or pruning rule that yields 1404 trials or identify any
random sampling or seed for this sweep.

The snapshot-level membership price is \(\varrho_{p,s}\) in
Eq.~\eqref{eq:membership-residual}; the paired quantity \(\varrho_{q,s}\) is
defined analogously. The group-preserving price is reported as the relative
change of the total represented field. The report states a \(1\%\) budget for
the quantity spent by each mode, but the available record does not specify
whether the snapshot and two-group residuals are combined by a maximum, mean,
quantile, or another reduction, nor does it expose the exact ensemble-level
denominator. It also does not state whether group covariances are explicitly
recentered and re-estimated after every trial, although total covariance in mode
A is unchanged algebraically.

For a trial with signed alignment vector
\(\boldsymbol\eta^R=\{\eta_{pq}^R\}_{p\le q}\), the recorded diagnostics are
\[
 d_{\mathrm{dom}}=\operatorname*{arg\,max}_{p\le q}|\eta_{pq}^R|,
 \qquad
 c_\eta=
 \frac{\boldsymbol\eta^R\!\cdot\boldsymbol\eta_0^R}
 {\|\boldsymbol\eta^R\|_2\|\boldsymbol\eta_0^R\|_2},
 \qquad
 \Delta\mathcal C_R=\mathcal C_R-\mathcal C_{R,0},
\]
together with the componentwise sign topology. A trial breaks the reported
attribution criterion if the dominant pair changes, any reported alignment sign
changes, or \(c_\eta<0.95\). The report names the largest admissible \(t\) for
each parameter tuple, but it does not state the global ordering objective or
tie rule used to select the single ``strongest admissible'' trial. The precise
normalization after the mode-B projections is likewise absent.

The intended counting protocol is therefore:
\begin{enumerate}
\item Freeze the declared representation, grouping, \(M_u\), ensemble, group
spans, principal system, and baseline centered covariance.
\item Retain pairwise directions at each \(\tau_{\mathrm{PA}}\), then construct
the snapshot-level exchanges for every declared
\((\alpha_{\mathrm{alloc}},t)\) and for modes A and B.
\item Recenter and recompute every object required by the trial-level stress
blocks, signed alignments, dominant pair, sign topology, \(c_\eta\), and
\(\Delta\mathcal C_R\), according to the declared covariance-update rule.
\item Reduce the snapshot- and group-level prices to the declared admissibility
quantity, apply the \(1\%\) gate, and count total, admitted, and
criterion-breaking trials.
\end{enumerate}
The aggregate record gives \(1404/509/37\) for those three counts, but steps
2--4 cannot reproduce them until the missing protocol details and trial-level
evidence are restored. These counts are not used to assign the Gaussian result
a binary verdict in this article.

\paragraph*{Integrity-basis coefficient control.}
The integrity-basis control is fully specified as follows. For each snapshot,
a spectral Gaussian filter of width \(2\Delta x\) uses the multiplier
\[
 \widehat G(\mathbf k)=
 \exp[-|\mathbf k|^2(2\Delta x)^2/24].
\]
The subfilter stress is
\(\tau_{ij,s}=\overline{u_{i,s}u_{j,s}}-\bar u_{i,s}\bar u_{j,s}\),
\(k_{\mathrm{sgs},s}=\tau_{ii,s}/2\), and the fit retains only points with
\(k_{\mathrm{sgs},s}>0\). The symmetric and antisymmetric parts of
\(\nabla\bar{\mathbf u}_s\) are both divided pointwise by the Frobenius norm of
the symmetric part. The ten Pope tensors \(\mathbf T_s^{(n)}\) then define
atoms
\[
 \mathbf B_s^{(n)}=2k_{\mathrm{sgs},s}\mathbf T_s^{(n)}
 \mathbb I[k_{\mathrm{sgs},s}>0],\qquad n=1,\ldots,10,
\]
which are fitted to the deviatoric subfilter stress with unit weights on the
three diagonal components and weight two on the three off-diagonal components.
Per-snapshot spatially global vectors are
recorded using ridge
\(10^{-12}\max\operatorname{diag}(G_{T,s})\). The reported geometry and shares
instead use the pooled Gram \(G_T=\sum_sG_{T,s}\), pooled right-hand side, and
one pooled spatially global vector \(\mathbf g\), fitted with ridge
\(10^{-12}\max\operatorname{diag}(G_T)\). No local invariant-dependent
coefficient field is fitted.

For coefficient perturbations, the \(f\)-budget is
\[
 \delta\mathbf g^{\mathsf T}G_T\delta\mathbf g
 \le f^2\,\mathbf g^{\mathsf T}G_T\mathbf g,
\]
so \(f=0.01\) is relative to the norm of the fitted represented stress, not to
the DNS stress. A pseudorandom generator initialized with seed \(20260801\)
draws 2048 independent vectors \(\mathbf z_0\sim\mathcal N(0,I_{10})\), sets
\[
 \mathbf z=\frac{\mathbf z_0}
 {(\mathbf z_0^{\mathsf T}G_T\mathbf z_0)^{1/2}},\qquad
 \mathbf g'=\mathbf g+0.01
 (\mathbf g^{\mathsf T}G_T\mathbf g)^{1/2}\mathbf z,
\]
and evaluates each draw once; no rejection or resampling is used.
The Gram matrix and basis fields remain fixed; only the coefficient allocation
and the normalized term and degree-group shares are recomputed. No covariance
re-estimation or local invariant-dependent coefficient field enters this
control. Separate worst-direction checks use both signs of the least-eigenvalue
direction at \(f=0.01\) and \(0.05\), retaining the sign that maximizes the
largest absolute term-share change. The 2048 draws leave \(T_8\) dominant at
\(1\%\); the \(5\%\) least-eigenvector check moves the maximum to \(T_2\).
The control records term and degree-group shares, cancellation indices,
dominant term, and maximum term-share change. It does not apply the Gaussian
membership gate or validate a trained closure network; the fitted-stress error
\(0.516\) bounds its interpretation.

An independent reconstruction of all 128 processed selection snapshots
reproduces the 2048 draws and dominant-term results to relative and absolute
tolerances \(2\times10^{-12}\). This check does not cover the unavailable
trial-level evidence behind the Gaussian \(1404/509/37\) counts.

\paragraph*{Snapshot-relative CVS control.}
For each of the 128 selection snapshots and each level-3 Coiflet-2,
Daubechies-4, and Symlet-4 construction, a periodized wavelet transform
is applied componentwise and one iterative Donoho threshold is computed from
the concatenated three-component vorticity coefficient vector. Starting from
\(T=(2Z\log N_c)^{1/2}\), with \(Z\) the mean squared coefficient and \(N_c\)
the concatenated coefficient count, \(Z\) is recomputed from coefficients with
\(|c|\le T\) for at most 12 iterations or until the retained coefficient count
stops changing. Coefficients with \(|c|>T\) are coherent.
Membership therefore changes with snapshot. Coherent and incoherent
coefficients are inverted separately and Biot--Savart decoded; the two group
fields are then centered over the ensemble. Principal cosines use the two
rank-127 realized-field ensemble spans, and the stress blocks are their direct
sample covariances. There is no fixed CVS group decoder or cross-snapshot
coefficient index whose covariance is propagated. Wavelet sensitivity compares
the three resulting half-share alignment vectors; their largest componentwise
range is \(0.0652\). No paired relabelling is run, so this is a wavelet-family
sensitivity observation rather than a binary verdict. The coefficient split
reconstructs vorticity to the reported numerical tolerance. The decoder omits
the velocity zero mode. The reconstruction gives a maximum-norm group-sum
discrepancy of \(0.288\)
relative to the maximum input velocity; the fitted-stress error is \(0.456\).

\subsection{Channel-flow protocol}

The channel extension represents \(\mathbf u'=\mathbf u-\mathbf U(y)\) on
sixteen \(64^3\) \(x\)--\(z\) blocks per time. Independently fitted wall and
outer regions occupy \(0\le j_y<64\) (\(y^+\simeq0\)--\(284\)) and
\(65\le j_y<129\) (\(y^+\simeq302\)--\(1002\)); plane \(64\) is excluded.
A \(K^3\) Gaussian lattice gives \(M_b=3K^3\) curl-lifted columns, with taper
\(1-\exp[-((y+1)/(15\delta_\nu))^2]\) at the wall and unity outside. Each block
fit uses \(73728\) quadrature-weighted points (seed \(20260723\)) and relative
ridge \(10^{-8}\).

Geometry determination uses six times and parameter selection
\(\{1600,3466\}\). The initial \(K20\) diagnostic uses
\(\{200,700,1200,1733,2266,2800,3333,3800\}\). After \(K23\) selection, the
frozen representation is evaluated once at
\(\{100,350,617,1000,1400,2200,2733,3650\}\), refitting only amplitudes.

Final \(K23\) uses \(N_G=12167\), \(M_b=36501\), update \(40\), and coherence
weight \(7297\). Its block-product objective and gradient match the dense
construction to relative error \(10^{-7}\) without materializing the full design;
the geometry and evidence roles are frozen before held-out evaluation.

The matched channel reference replaces each Gaussian vector potential by
\[
  \mathbf A^{B}_{\ell,c}
  =
  w(d_w)B_{\ell_x}^{(3)}(x)B_{\ell_y}^{(3)}(y)
  B_{\ell_z}^{(3)}(z)\mathbf e_c,
  \qquad
  \boldsymbol\phi^{B}_{\ell,c}
  =
  \nabla\times\mathbf A^{B}_{\ell,c},
\]
using deterministic clamped-knot splines. Rank-revealing selection on the
geometry and selection data retains \(M_b=36501\) directions before the
held-out split is read; taper, curl lift, sampling, ridge, precision, and
held-out times otherwise match \(K23\).

Production uses frozen nonuniform-\(y\) derivatives and quadrature. The
99th-percentile normalized common-face jumps span \(0.743\)--\(1.461\) for
velocity and \(0.916\)--\(1.411\) for stress, excluding a globally continuous
block decoder. Face trimming excludes interface cells as the source of the
reported wall-normal stress errors. The supported scope is therefore
boundary-compatible stress representation and attribution in independently
decoded local regions.

\begin{table}
  \centering
  \caption{\label{tab:channel-fidelity}
  Channel-flow stress fidelity.}
  \footnotesize
  \setlength{\tabcolsep}{3.5pt}
  \begin{ruledtabular}
  \begin{tabular}{llrrrr}
    Representation\footnote{Reported frozen-model values use eight held-out
    times. The matched B-spline protocol is specified in the surrounding text.}
    & Region & \(M_b\)
    & \(e_R\)\footnote{\(K23\) and B-spline use the held-out times; \(K26\) is
    a parameter-selection diagnostic, and -- denotes an unretained value.}
    & \(e_{uv}\)
    & \(\kappa(K)\)\footnote{Condition number of the normalized positive
    Gram spectrum; admissibility requires \(\kappa(K)<10^8\).} \\
    \colrule
    frozen \(K23\) & outer & \(36501\)
    & \(0.0661\) & \(0.0269\) & \(9.52\times10^6\) \\
    frozen \(K23\) & wall & \(36501\)
    & \(0.129\) & \(0.123\) & \(3.04\times10^7\) \\
    \(K26\) candidate & outer & \(52728\)
    & \(\simeq0.043\) & -- & \(\simeq1.3\times10^8\) \\
    \(K26\) candidate & wall & \(52728\)
    & \(\simeq0.078\) & -- & \(\simeq2.4\times10^8\) \\
    matched B-spline & outer & \(36501\)
    & \(0.401\) & \(0.417\) & \(1.18\times10^3\) \\
    matched B-spline & wall & \(36501\)
    & \(0.440\) & \(0.565\) & \(8.56\times10^2\) \\
  \end{tabular}
  \end{ruledtabular}
\end{table}

\section{Principal-direction fidelity}
\label{app:stress-orientation}

\begingroup
The represented stress preserves its principal directions as well as its
eigenvalue-based componentality over the resolved T202 interior. We evaluate
\(R^{\mathrm{dep}}\) on the trim-four domain
\(\Omega_{\mathrm{int}}\) used in
Sec.~\ref{subsec:jhtdb-stress-state}.
\endgroup
For ordered eigenpairs
\((\lambda_i,\mathbf v_i)\) of the anisotropy tensor \(b\), define
\[
  \theta_1
  =
  \arccos\left|
    \mathbf v_1^{\mathrm{DNS}}\!\cdot\mathbf v_1^{\mathrm{dep}}
  \right|,
  \qquad
  \gamma_1
  =
  \frac{\lambda_1^{\mathrm{DNS}}-\lambda_2^{\mathrm{DNS}}}
       {\|b^{\mathrm{DNS}}\|_F+10^{-12}} .
\]
The leading axis is evaluated where
\(k^{\mathrm{DNS}}>0.1\langle k^{\mathrm{DNS}}\rangle_\Omega\),
\(\|b^{\mathrm{DNS}}\|_F>0.05\), and
\(\gamma_1>\tau_{\mathrm{gap}}\). The principal value is
\(\tau_{\mathrm{gap}}=0.15\), with \(0.10\) and \(0.20\) used for
sensitivity. The complete normalized deviatoric state is compared through
\[
  \theta_b
  =
  \arccos\!\left[
    \operatorname{clip}\!\left(
      \frac{b^{\mathrm{DNS}}:b^{\mathrm{dep}}}
           {\|b^{\mathrm{DNS}}\|_F\|b^{\mathrm{dep}}\|_F+10^{-12}},
      -1,1
    \right)
  \right],
\]
where both anisotropy norms exceed \(0.05\).

\begin{figure}
  \centering
  \includegraphics[width=\linewidth]
  {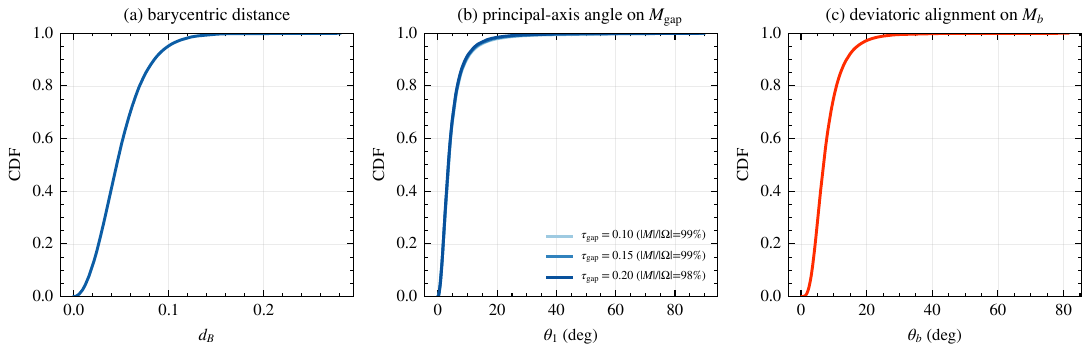}
  \caption{Tensor-state fidelity on the T202 trim-four interior:
  (a) barycentric distance \(d_B\), (b) leading-axis error \(\theta_1\) for
  three eigengap thresholds, and (c) deviatoric-tensor angle \(\theta_b\).}
  \label{fig:app-orientation-cdf}
\end{figure}

At \(\tau_{\mathrm{gap}}=0.15\), the median and 90th-percentile
\(\theta_1\) are \(3.50^\circ\) and \(9.47^\circ\), with \(98.75\%\)
coverage. Sweeping \(\tau_{\mathrm{gap}}\) from \(0.10\) to \(0.20\)
changes the median from \(3.52^\circ\) to \(3.48^\circ\) and the 90th
percentile from \(9.66^\circ\) to \(9.25^\circ\).
The corresponding \(\theta_b\) median and 90th percentile are
\(6.83^\circ\) and \(13.85^\circ\)
[Fig.~\ref{fig:app-orientation-cdf}].

\begin{figure}
  \centering
  \includegraphics[width=\linewidth]
  {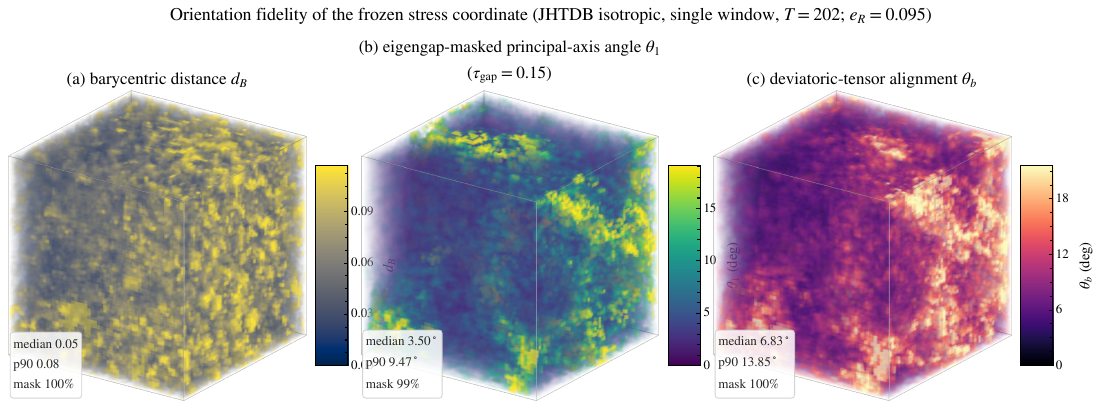}
  \caption{Spatial support of T202 tensor-state residuals:
  (a) \(d_B\), (b) \(\theta_1\) at \(\tau_{\mathrm{gap}}=0.15\), and
  (c) \(\theta_b\).
  Angular errors are in degrees.}
  \label{fig:app-orientation-spatial}
\end{figure}

The upper-tail errors are localized rather than volume-filling
(Fig.~\ref{fig:app-orientation-spatial}). This conclusion is restricted to the
trimmed interior: \(e_R\) is \(0.095\) there and \(0.150\) on the full grid,
where the nonperiodic cutout meets the periodic decoder.

\section{Spatial organization of covariance cancellation}
\label{app:spatial-attribution-slices}

\begin{figure}
  \centering
  \includegraphics[width=\linewidth]
  {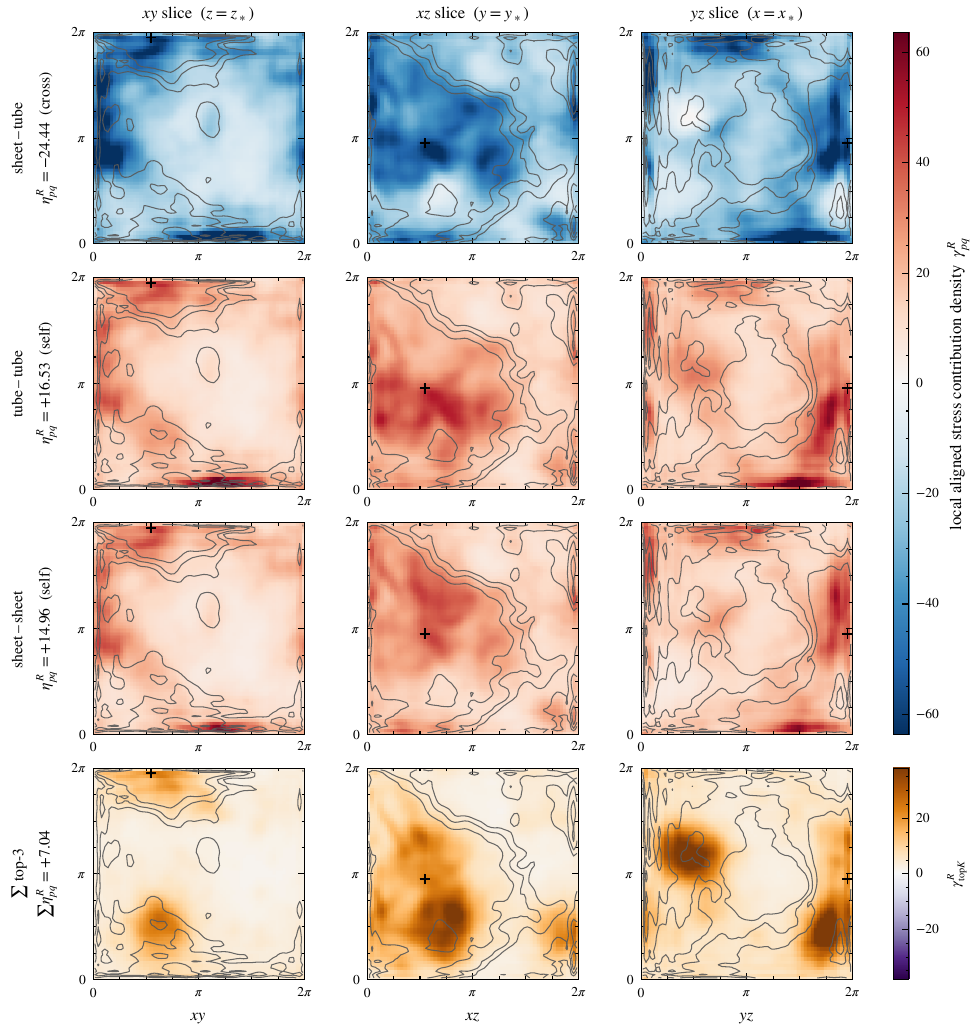}
  \caption{Slices of the dominant sheet--tube, tube--tube, and sheet--sheet
  covariance blocks and their sum on T202. Red aligns with
  \(R^{\mathrm{coef}}\), blue opposes it, and gray contours show
  \(\|R^{\mathrm{coef}}\|_F\).}
  \label{fig:app-jhtdb-spatial-attribution}
\end{figure}

\begingroup
The dominant self and cross blocks cancel within the same spatial regions.
Figure~\ref{fig:app-jhtdb-spatial-attribution} compares the sheet--tube cross
block with the tube--tube and sheet--sheet self blocks on slices through
\((17,62,30)\). Their similar patterns and opposite signs are consistent with
the full-volume self--cross correlation \(r=-0.9993\) in
Fig.~\ref{fig:jhtdb-covariance-closure}.
\endgroup

\begingroup
The displayed blocks sum to \(+7.05\), the remaining blocks to \(-6.05\),
and all six to unity. The negative cross covariance is therefore a spatially
distributed counterfield to the positive self covariance, rather than an
integral cancellation between disjoint regions.
\endgroup

\clearpage

\section[Channel-flow tensor fidelity]
{Channel-flow tensor fidelity}
\label{app:channel-tensor-state}

\begingroup
The frozen channel representation preserves stress directions in both local
regions, while stress magnitude and componentality are less accurate near the
wall. We evaluate the full three-dimensional stress on the eight held-out
times of Sec.~\ref{subsec:channel-fidelity} and sixteen homogeneous
\(x\)--\(z\) translations per time. Wall and outer regions are decoded
independently. The analysis uses the four-cell-trimmed \(x\)--\(z\) interior,
nonuniform-\(y\) quadrature, and the diagnostics of
Appendix~\ref{app:stress-orientation}.
\endgroup

Wall-region weighted
medians and 90th percentiles are \(0.059\) and \(0.109\) (\(d_B\)),
\(3.08^\circ\) and \(6.55^\circ\) (\(\theta_1\)), and
\(6.66^\circ\) and \(11.78^\circ\) (\(\theta_b\)).  The corresponding outer
values are \(0.042\) and \(0.078\), \(2.96^\circ\) and \(6.37^\circ\), and
\(6.57^\circ\) and \(12.26^\circ\).  The displayed masks retain
\(76\%\)--\(77\%\) of the full-block quadrature weight, set mainly by the
prescribed \(x\)--\(z\) trim at \(76.6\%\) of the block.

\begin{figure*}
  \centering
  \includegraphics[width=\textwidth]
  {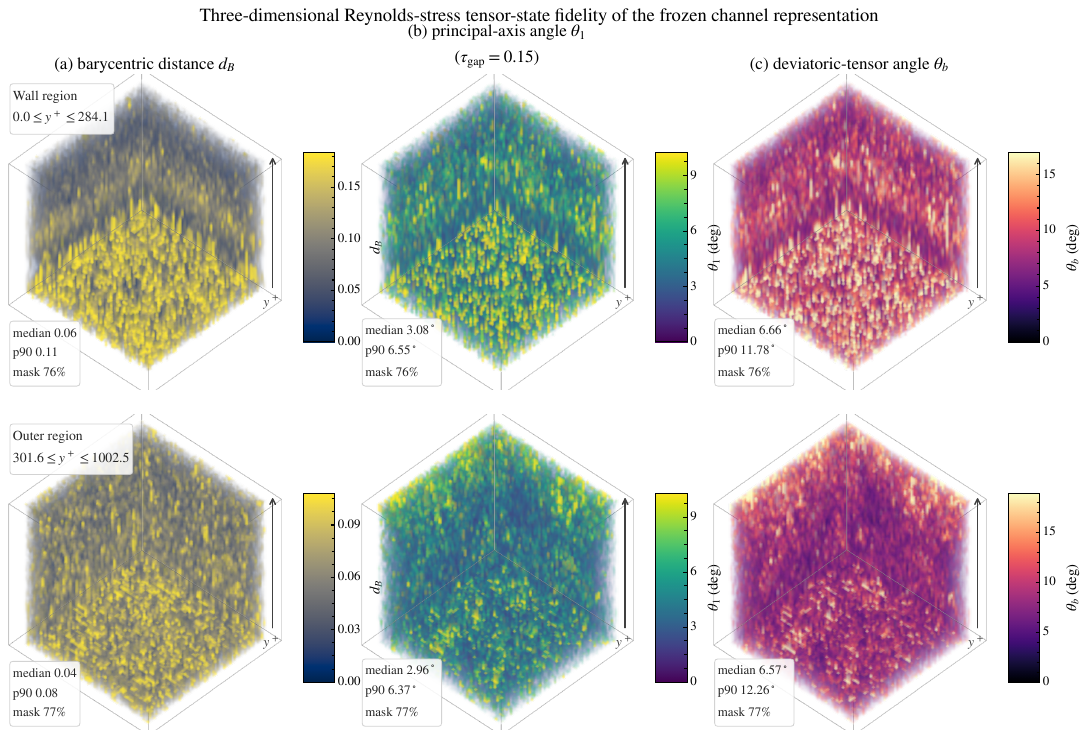}
  \caption{Local channel-flow tensor fidelity. Rows show the independently
  decoded wall and outer regions; columns show (a) barycentric distance
  \(d_B\), (b) principal-axis angle \(\theta_1\) at
  \(\tau_{\mathrm{gap}}=0.15\), and (c) deviatoric-tensor angle \(\theta_b\).
  Transparent voxels fail the diagnostic masks. The two regions use separate
  local coordinates and are not stitched.}
  \label{fig:app-channel-tensor-state}
\end{figure*}

\begingroup
The three-dimensional tensor errors are \(0.185\) (wall) and
\(0.101\) (outer), compared with profile errors of \(0.129\) and \(0.0661\).
The increase is expected because the
three-dimensional norm retains local \(x\)--\(z\) covariance variation that
can cancel under planar averaging.  Averaging the volumetric stresses
over \(x\) and \(z\) recovers every reported profile component to
within \(5\times10^{-11}\).  The wall-to-outer fidelity contrast is therefore
primarily associated with stress magnitude, componentality, and wall-normal
localization rather than a systematic rotation of the stress axes.
\endgroup

\clearpage
\bibliography{aipsamp}

\end{document}


\title{Supplementary material for\\
``Identifiability Limits of Reynolds-Stress Attribution in Localized
Nonorthogonal Flow Representations''}

\author{Peng Cheng}
 \email{peng.cheng.hit@gmail.com}
\author{Ming Jian}%
\author{Wen Tong}
\affiliation{Huawei Technologies Canada Co., Ltd., \\ 303 Terry Fox Drive, Suite 101, Ottawa, ON, K2K 3J1, Canada}

\date{\today}

\maketitle

\noindent
\begingroup
This document supports the physical argument of the main article with the
nomenclature (Sec.~\ref{supp:nomenclature}), the matched represented-variable
control (Sec.~\ref{supp:represented-variable-ablation}), and the evidence scope
(Sec.~\ref{supp:audit-protocol}). Numbers without an
\(\mathrm S\) prefix refer to the main article.
\endgroup

\section{Nomenclature}
\label{supp:nomenclature}

\begingroup
The notation is grouped by geometry, fields, diagnostics, and perturbations;
the final column points to each definition. A hat denotes a represented
quantity. Superscripts \(\mathrm{coef}\) and \(\mathrm{dep}\) distinguish the
Gaussian channel from the full field containing the exact snapshot-wise
constant channel. Signed pair alignments sum to one but are not nonnegative
shares. Envelope parameters use \(\theta\), span angles use \(\vartheta\), and
the Kolmogorov length is \(\eta_K\).
\endgroup

\begin{table}
\begingroup
\caption{\label{tab:notation}
Geometry and dictionary notation.}
\footnotesize
\setlength{\tabcolsep}{4pt}
\renewcommand{\arraystretch}{1.15}
\begin{tabular}{@{}lp{0.545\textwidth}l@{}}
\toprule
Symbol & Definition & Defined in \\
\midrule
\(\Omega=\mathbb T^3\), \(\nabla_h\), \(\mathcal S_h\)
& Periodic analysis domain; discrete spectral curl and gradient; frozen
deterministic grid sampler.
& Sec.~\ref{M-subsec:vorticity-decoding}, Sec.~\ref{M-subsec:graph-norm-fitting}\\
\(\boldsymbol\mu_i\), \(\Sigma_i\)
& Prescribed envelope center, held fixed; envelope covariance in fixed
isotropic, learned isotropic, axis-aligned, or freely oriented form.
& Eq.~\eqref{M-eq:gaussian-geometry}\\
\(\theta\), \(\theta^\star\)
& Learnable envelope geometry (widths and, for oriented families, rotations);
its value at the geometry-determination minimum.
& Eq.~\eqref{M-eq:gaussian-geometry}, Eq.~\eqref{M-eq:learning-objective}\\
\(g_i^\theta\)
& Scalar Gaussian envelope sampled at center \(i\); it describes support
geometry rather than a vector field.
& Eq.~\eqref{M-eq:gaussian-geometry}\\
\(\mathbf g_{i,c}^\theta\), \(\mathcal N_W\)
& Curl-lifted solenoidal vorticity atom from envelope \(i\) and Cartesian seed
\(\mathbf e_c\); normalization in the sampled fitting metric.
& Eq.~\eqref{M-eq:curl-lifted-atoms}\\
\(G_\theta\), \(G=G_{\theta^\star}\)
& Gaussian vorticity dictionary (synthesis operator), discretized as
\(N_\omega\times M_b\); its frozen form.
& Eq.~\eqref{M-eq:curl-lifted-atoms}, Sec.~\ref{M-subsec:graph-norm-fitting}\\
\(\mathcal V_{\omega,h}^{0,\mathrm{div}}\)
& Discrete zero-mean solenoidal vorticity space containing the range of
\(G_\theta\).
& Eq.~\eqref{M-eq:curl-lifted-atoms}\\
\(\mathcal K\)
& Fixed periodic Biot--Savart decoder returning zero-mean, discretely
divergence-free velocity.
& Eq.~\eqref{M-eq:biot-savart-decoder}\\
\(\Phi=\mathcal K G\), \(\Phi_p\)
& Frozen decoded velocity dictionary; its column block for group \(p\).
& Eq.~\eqref{M-eq:representation-decoder},
Eq.~\eqref{M-eq:group-pair-accounting}\\
\(\mathbf k\), \(\mathbf k_d\)
& Fourier wavevector; Nyquist-zeroed derivative wavevector used for the
snapshot curl and the Gaussian curl lift.
& Eq.~\eqref{M-eq:biot-savart-decoder}\\
\(N_G\), \(M_b=3N_G\)
& Number of Gaussian envelopes; number of dictionary columns and of amplitudes
fitted per snapshot.
& Eq.~\eqref{M-eq:curl-lifted-atoms}\\
\(W\), \(H_\ell\), \(\ell\)
& Sampled fitting metric \(H_\ell^{\mathsf T}\mathcal S_h^{\mathsf T}\mathcal
S_hH_\ell\); Fourier multiplier \(w_\ell^{1/2}\); graph-norm length in
\(w_\ell(\mathbf k)=|\mathbf k|^{-2}+\ell^2\).
& Eq.~\eqref{M-eq:amplitude-fit}, Eq.~\eqref{M-eq:graph-norm}\\
\(\lambda_{\mathrm{ridge}}\)
& Ridge parameter of the amplitude solve.
& Eq.~\eqref{M-eq:amplitude-fit}\\
\(\beta_{\mathrm{stress}}\), \(\beta_{\mathrm{spectral}}\)
& Stress-aware geometry-objective weight, set to zero in the retained
three-dimensional runs; spectral-error weight in the two-dimensional
represented-variable control.
& App.~\ref{M-app:reproducibility},
Sec.~\ref{supp:represented-variable-ablation}\\
\(\mathcal R_{\mathrm{geom}}\), \(\mathcal R_{\mathrm{Gram}}\)
& Indicator of the admissible width--aspect set; thresholded coherence penalty
on the normalized Gram correlations.
& Eq.~\eqref{M-eq:learning-objective}, Eq.~\eqref{M-eq:gram-regularization}\\
\(\alpha\), \(\tau_{\mathrm{Gram}}\)
& Weight and correlation threshold of \(\mathcal R_{\mathrm{Gram}}\); the
threshold permits local overlap without imposing global orthogonality.
& Eq.~\eqref{M-eq:gram-regularization}\\
\(K_\theta\), \(K_G\); \(C_\theta\), \(C_G\)
& Graph-metric Gram matrix \(G_\theta^{\mathsf T}WG_\theta\) and its normalized
correlation matrix; their frozen values at \(\theta^\star\).
& Eq.~\eqref{M-eq:gram-regularization},
Sec.~\ref{M-subsec:graph-norm-fitting}\\
\(\kappa(K_G)\), \(r_{\mathrm{eff}}(K_G)\), \(E_{\mathrm{coh}}\)
& Condition number, effective rank, and residual coherence
\(\operatorname*{mean}_{m<n}[\max(0,|(C_G)_{mn}|-\tau_{\mathrm{Gram}})]^2\) of
the frozen Gram.
& Table~\ref{M-tab:coord-verify}\\
\(\rho_{\max}\)
& Aspect-ratio bound imposed on the envelopes during geometry determination.
& Sec.~\ref{M-subsec:jhtdb-group-attribution}, App.~\ref{M-app:reproducibility}\\
\(N_{\mathrm{geom}}\), \(N_{\mathrm{opt}}\), \(B\)
& Geometry-determination ensemble size, optimizer-update count, and their
product \(B=N_{\mathrm{geom}}N_{\mathrm{opt}}\), used only as an exposure
proxy.
& Sec.~\ref{M-subsec:verify-geometry-selection}\\
\(\mathbf A_{i,c}\), \(\boldsymbol\phi^{\,w}_{i,c}\)
& Channel vector potential; its curl, which is a column of the channel
velocity dictionary.
& Eq.~\eqref{M-eq:channel-wall-representation}\\
\(w(d_w)\), \(\ell_w\), \(\delta_\nu\)
& Wall taper \(1-\exp[-(d_w/\ell_w)^2]\) at wall distance \(d_w\); taper length
\(\ell_w=15\delta_\nu\); viscous length scale.
& Eq.~\eqref{M-eq:channel-wall-representation}\\
\(K\)
& Channel Gaussian lattice size; a \(K^3\) lattice generates
\(M_b=3K^3\) vector-potential curl columns.
& Sec.~\ref{M-subsec:channel-coordinate}, App.~\ref{M-app:reproducibility}\\
\bottomrule
\end{tabular}
\endgroup
\end{table}

\begin{table}
\begingroup
\caption{\label{tab:notation-fields}
Field and ensemble notation.}
\footnotesize
\setlength{\tabcolsep}{4pt}
\renewcommand{\arraystretch}{1.15}
\begin{tabular}{@{}lp{0.545\textwidth}l@{}}
\toprule
Symbol & Definition & Defined in \\
\midrule
\(S\)
& Number of snapshots in the evaluated ensemble.
& Sec.~\ref{M-subsec:vorticity-decoding}\\
\(U\), \(Y\)
& DNS velocity and vorticity ensembles, \(U\in\mathbb R^{S\times N_u}\) and
\(Y\in\mathbb R^{S\times N_\omega}\); rows are vectorized snapshots.
& Sec.~\ref{M-subsec:vorticity-decoding}\\
\(A\), \(A'=P_SA\)
& Snapshot-dependent amplitudes; their ensemble-centered form, with
\(P_S=\mathbf I_S-S^{-1}\mathbf 1_S\mathbf 1_S^{\mathsf T}\).
& Eq.~\eqref{M-eq:representation-decoder}, Eq.~\eqref{M-eq:covariance-lift}\\
\(C_A\), \(C_{pq}\)
& Amplitude covariance \((S-1)^{-1}A'^{\mathsf T}A'\); its block for groups
\(p\) and \(q\).
& Eq.~\eqref{M-eq:covariance-lift},
Eq.~\eqref{M-eq:group-pair-accounting}\\
\(\widehat Y\), \(\widehat U^{G}\)
& Represented vorticity \(AG^{\mathsf T}\); Gaussian-channel represented
velocity \(A\Phi^{\mathsf T}\).
& Eq.~\eqref{M-eq:representation-decoder}\\
\(\overline U_h\), \(\widehat U\)
& Snapshot-wise spatial means \(\langle\mathbf u_s\rangle_h\) broadcast over
the grid; full represented ensemble
\(\widehat U=\overline U_h+\widehat U^{G}\).
& Eq.~\eqref{M-eq:hodge-deployment}\\
\(R^{\mathrm{coef}}\)
& Gaussian-channel represented Reynolds stress \(\Phi C_A\Phi^{\mathsf T}\),
equal to the covariance of the coefficient-decoded velocity.
& Eq.~\eqref{M-eq:covariance-lift}\\
\(R^{\mathrm{dep}}\)
& Full represented stress \(\operatorname{Cov}_s(\widehat{\mathbf u}_s)\),
including the constant channel and both constant--Gaussian cross terms.
& Eq.~\eqref{M-eq:deployed-diagnostics}\\
\(R^{\mathrm{DNS}}\)
& Reference DNS sample stress \(\operatorname{Cov}(\mathbf u')\) of the same
ensemble.
& Table~\ref{M-tab:jhtdb-transfer}\\
\(k^X\), \(q^X\)
& For \(X\in\{\mathrm{coef},\mathrm{dep}\}\), the turbulent kinetic energy
\(k^X=\tfrac12\operatorname{tr}R^X\) and the stress divergence
\(q^X=-\nabla_h\cdot R^X\).
& Sec.~\ref{M-subsec:covariance-lift}\\
\(b\), \(\mathbf y_B\)
& Anisotropy tensor \(R/(2k)-\mathbf I/3\); barycentric coordinates determined
by its eigenvalues.
& Sec.~\ref{M-subsec:jhtdb-stress-state}\\
\(\Delta x\), \(\eta_K\), \(\mathrm{Re}_\lambda\)
& Spacing of the \(64^3\) analysis grid; Kolmogorov length of the source DNS
field; its Taylor-microscale Reynolds number.
& Sec.~\ref{M-subsec:vorticity-decoding},
App.~\ref{M-app:reproducibility}\\
\(\Omega_{\mathrm{int}}\)
& Trim-four \(56^3\) interior of the \(64^3\) analysis grid, used for the
orientation diagnostic.
& App.~\ref{M-app:stress-orientation}\\
T202
& Ensemble of \(202\) consecutive snapshots at one origin used in neither
geometry determination nor parameter selection.
& App.~\ref{M-app:reproducibility}\\
\(\tau_{\mathrm{corr,max}}\), \(L^\star\)
& Largest estimated temporal correlation time in the T202 dependence analysis;
selected circular-block length
\(L^\star=\lceil2\tau_{\mathrm{corr,max}}\rceil\).
& App.~\ref{M-app:reproducibility}\\
\(\mathbf u'_s\), \(\mathbf U(y)\), \(R_{ij}(y)\)
& Channel velocity fluctuation \(\mathbf u_s-\mathbf U(y)\); prescribed mean
profile; wall-normal stress profile
\(\langle u_i'u_j'\rangle_{s,x,z\mid y}\).
& Sec.~\ref{M-subsec:channel-coordinate},
Sec.~\ref{M-subsec:channel-fidelity}\\
\bottomrule
\end{tabular}
\endgroup
\end{table}

\begin{table}
\begingroup
\caption{\label{tab:notation-diagnostics}
Fidelity and attribution diagnostics.}
\footnotesize
\setlength{\tabcolsep}{4pt}
\renewcommand{\arraystretch}{1.15}
\begin{tabular}{@{}lp{0.545\textwidth}l@{}}
\toprule
Symbol & Definition & Defined in \\
\midrule
\multicolumn{3}{l}{\emph{Error and fidelity metrics}}\\
\(e_u\)
& Full-field relative velocity error
\(\|\widehat U-U\|_F/\|U\|_F\).
& Table~\ref{M-tab:coord-verify}\\
\(e_u^{\mathrm{med}}\)
& Median over snapshots of
\(\|\widehat{\mathbf u}'_s-\mathbf u'_s\|_2/\|\mathbf u'_s\|_2\).
& Table~\ref{M-tab:jhtdb-transfer}\\
\(e_{\mathrm{graph}}\)
& Relative error of the represented vorticity in the graph norm.
& Table~\ref{M-tab:coord-verify},
Eq.~\eqref{M-eq:graph-norm}\\
\(e_R\), \(e_R^{\mathrm{T202}}\)
& Full-field relative Reynolds-stress error
\(\|R^{\mathrm{dep}}-R^{\mathrm{DNS}}\|_F/\|R^{\mathrm{DNS}}\|_F\); the same
error evaluated on the \(S=202\) temporal ensemble.
& Table~\ref{M-tab:jhtdb-transfer},
Table~\ref{M-tab:coord-verify}\\
\(e_R^{\mathrm{coef}}\)
& The same error for the Gaussian channel alone,
\(\|R^{\mathrm{coef}}-R^{\mathrm{DNS}}\|_F/\|R^{\mathrm{DNS}}\|_F\), which is
the tensor decomposed by the group analysis.
& Table~\ref{M-tab:jhtdb-transfer}\\
\(e_k\), \(\chi_k\)
& Relative error of the represented TKE field; ratio of integrated represented
to integrated DNS TKE.
& Sec.~\ref{M-subsec:jhtdb-stress-state},
Table~\ref{M-tab:jhtdb-transfer}\\
\(e_q\)
& Relative stress-divergence error
\(\|q^{\mathrm{dep}}-q^{\mathrm{DNS}}\|_2/\|q^{\mathrm{DNS}}\|_2\).
& Table~\ref{M-tab:jhtdb-force-scope}\\
\(f_q^{B}\)
& Fraction of the squared stress-divergence residual falling in wavenumber
band \(B\in\{\mathrm L,\mathrm M,\mathrm H\}\); \(f_q^{\mathrm H}\) is the
high-wavenumber share.
& Table~\ref{M-tab:jhtdb-force-scope}\\
\(E_\infty\)
& Extrapolated stress-divergence-error floor, fitted from nested T202
prefixes.
& Table~\ref{M-tab:jhtdb-force-scope}\\
\(e_q^{H^{-1}}\)
& Stress-divergence error measured in the \(H^{-1}\) norm.
& Sec.~\ref{M-sec:spectral-enrichment}\\
\(d_B\)
& Barycentric distance
\(\|\mathbf y_B^{\mathrm{dep}}-\mathbf y_B^{\mathrm{DNS}}\|_2\), an
eigenvalue-based componentality error.
& Sec.~\ref{M-subsec:jhtdb-stress-state}\\
\(\theta_1\)
& Sign-invariant leading principal-axis error
\(\arccos|\mathbf v_1^{\mathrm{DNS}}\!\cdot\mathbf v_1^{\mathrm{dep}}|\).
& App.~\ref{M-app:stress-orientation}\\
\(\theta_b\)
& Angle between the normalized deviatoric tensors \(b^{\mathrm{DNS}}\) and
\(b^{\mathrm{dep}}\).
& App.~\ref{M-app:stress-orientation}\\
\(\gamma_1\), \(\tau_{\mathrm{gap}}\)
& Normalized DNS eigengap
\((\lambda_1^{\mathrm{DNS}}-\lambda_2^{\mathrm{DNS}})/
(\|b^{\mathrm{DNS}}\|_F+\varepsilon)\); the mask threshold on \(\gamma_1\)
below which \(\theta_1\) is not evaluated.
& App.~\ref{M-app:stress-orientation}\\
\(e_{uv}\), \(e_P\)
& Relative errors of the channel shear component of \(R\) and of the
production \(P=-R_{uv}\,\mathrm dU/\mathrm dy\).
& Sec.~\ref{M-subsec:channel-fidelity},
Sec.~\ref{M-subsec:channel-production-interface}\\
\(\epsilon_R\)
& Fidelity-resolution width; a stress-error difference below \(\epsilon_R\) is
not treated as a resolved gain.
& Sec.~\ref{M-subsec:verify-geometry-selection}\\
\(\mathcal S_{\mathrm{cov}}\), \(\mathcal B_{\mathrm{stab}}\)
& Empirically covariance-stable candidate set; its subset within
\(\epsilon_R\) of the minimum \(e_R\), from which the retained state is taken.
& Sec.~\ref{M-subsec:verify-geometry-selection}\\
\(\xi_{40}\)
& Cosine similarity of the signed closed blob--sheet--tube share vector
\(\boldsymbol\eta_{N_{\mathrm{opt}}}\) to the state at
\(N_{\mathrm{opt}}=40\).
& Sec.~\ref{M-subsec:verify-geometry-selection}\\
\midrule
\multicolumn{3}{l}{\emph{Attribution and identifiability diagnostics}}\\
\(\{\mathcal G_p\}_{p=1}^{P_g}\)
& Partition of the frozen Gaussian columns into \(P_g\) groups; each envelope
label is inherited by its three curl channels.
& Sec.~\ref{M-subsec:group-attribution}\\
\(R^{(p,q)}\), \(\widetilde R^{(p,q)}\)
& Ordered pair contribution
\(\Phi_pC_{pq}\Phi_q^{\mathsf T}\); its nonduplicated symmetric form, with
\(R^{\mathrm{coef}}=\sum_{p\le q}\widetilde R^{(p,q)}\).
& Eq.~\eqref{M-eq:group-pair-accounting},
Sec.~\ref{M-subsec:group-attribution}\\
\(R^{[p]}\), \(k^{[p]}\), \(q^{[p]}\)
& Closed group stress under the half-share convention, and its trace and
divergence diagnostics; they sum to the corresponding total.
& Eq.~\eqref{M-eq:closed-group-accounting}\\
\(\langle\cdot,\cdot\rangle_{L_F^2}\)
& Domain-integrated tensor Frobenius inner product.
& Eq.~\eqref{M-eq:pair-alignment}\\
\(\eta_{pq}^{R}\), \(\boldsymbol\eta^R\)
& Signed alignment of \(\widetilde R^{(p,q)}\) with \(R^{\mathrm{coef}}\),
summing to one; the vector \(\{\eta_{pq}^R\}_{p\le q}\) tracked by the
identifiability tests.
& Eq.~\eqref{M-eq:pair-alignment},
Sec.~\ref{M-subsec:group-attribution}\\
\(\mathcal C_R\)
& Cancellation index \(\sum_{p\le q}|\eta_{pq}^{R}|\), measuring the
opposition hidden by the unit sum of the \(\eta_{pq}^{R}\).
& Eq.~\eqref{M-eq:pair-alignment}\\
\(r_{\mathrm{SC}}\)
& Correlation between the summed self and summed cross alignment densities
over the analysis volume.
& Sec.~\ref{M-subsec:jhtdb-group-attribution}\\
\(\eta_{uv}^{\mathrm{self}}\), \(\eta_{uv}^{\mathrm{cross}}\),
\(\zeta_{uv}\)
& Integrated self and cross shear alignments normalized by the net represented
shear, and their cancellation fraction
\(|\eta_{uv}^{\mathrm{cross}}|/\eta_{uv}^{\mathrm{self}}\).
& Sec.~\ref{M-subsec:channel-attribution}\\
\(M_u\)
& Full-grid kinetic-energy metric used for span comparison.
& Eq.~\eqref{M-eq:span-identifiability}\\
\(\mathcal S_p\), \(Q_p\)
& Decoded velocity span \(\operatorname{range}(\Phi_p)\) of group \(p\); a
basis of it orthonormal in \(M_u\).
& Eq.~\eqref{M-eq:span-identifiability}\\
\(\cos\vartheta_j^{(p,q)}\)
& Principal cosines \(\sigma_j(Q_p^{\mathsf T}M_uQ_q)\) between the decoded
spans of groups \(p\) and \(q\); values near unity mark near-shared
directions.
& Eq.~\eqref{M-eq:span-identifiability}\\
\bottomrule
\end{tabular}
\endgroup
\end{table}

\begin{table}
\begingroup
\caption{\label{tab:notation-perturbations}
Perturbation notation.}
\footnotesize
\setlength{\tabcolsep}{4pt}
\renewcommand{\arraystretch}{1.15}
\begin{tabular}{@{}lp{0.545\textwidth}l@{}}
\toprule
Symbol & Definition & Defined in \\
\midrule
\(\tau_{\mathrm{PA}}\), \(n_{\mathrm{PA}}\)
& Principal-cosine threshold used by the relabelling tests; number of retained
near-shared directions for a group pair.
& Eq.~\eqref{M-eq:exchange-basis}\\
\(\mathbf u_{p,s}\), \(\pi_{j,s}^{p}\), \(T_{j,s}\)
& Snapshot-level field assigned to group \(p\); its coefficient along the
principal bisector \(\mathbf d_j\); total exchangeable coefficient for the
selected group pair.
& Eq.~\eqref{M-eq:field-preserving-relabelling}\\
\(\alpha_{\mathrm{alloc}}\), \(t\), \(\delta_{j,s}\)
& Allocation fraction, perturbation magnitude, and snapshot-level exchange
coefficient of the declared relabelling family.
& Eq.~\eqref{M-eq:field-preserving-relabelling}\\
\(\varrho_{p,s}\), \(\varrho_{q,s}\)
& Snapshot-level relative departures of the two relabelled group fields from
their respective declared spans.
& Eq.~\eqref{M-eq:membership-residual}\\
\(G_T\), \(\mathbf g\), \(f\)
& Pooled ten-term integrity-basis Gram matrix; pooled spatially global
coefficient vector; relative represented-stress-norm perturbation budget.
& App.~\ref{M-app:relabel-protocol}\\
\(\mathbf z_0\), \(\mathbf z\)
& Standard-normal draw in the ten-dimensional coefficient space; the same draw
normalized to unit \(G_T\)-norm for the integrity-basis control.
& App.~\ref{M-app:relabel-protocol}\\
\bottomrule
\end{tabular}
\endgroup
\end{table}

\clearpage

\section{Represented-variable control}
\label{supp:represented-variable-ablation}

At fixed Gaussian geometry and storage, vorticity gives the lowest
velocity, vorticity, and velocity-gradient errors across the tested budgets.
The matched two-dimensional comparison uses normalized Gaussian blending
\cite{shenoy2026gaussianfield} and additive Gaussian velocity
\cite{xing2025gaussian}; scalar controls represent streamfunction or
vorticity:
\begin{align*}
  \mathbf u_G^{\rm norm}(\mathbf x)
    &=\sum_i w_i(\mathbf x)\mathbf a_i,
    &w_i(\mathbf x)
    &=\frac{G_i(\mathbf x)}
      {\sum_jG_j(\mathbf x)+\epsilon},\\
  \mathbf u_G^{\rm add}(\mathbf x)
    &=\sum_i G_i(\mathbf x)\mathbf a_i,\\
  \psi_G(\mathbf x)
    &=\sum_i a_iG_i(\mathbf x),\\
  \omega_G(\mathbf x)
    &=\sum_i a_iG_i(\mathbf x).
\end{align*}
The streamfunction control sets
\(\mathbf u_G^\psi=\nabla^\perp\psi_G\); the vorticity control solves
\(-\Delta\psi_G^\omega=\omega_G\) at zero mean, giving
\(\mathbf u_G^\omega=\nabla^\perp\psi_G^\omega\)
\cite{majda2002vorticity}, the scalar analogue of the three-dimensional
curl-lifted construction.  Direct velocity maps store \(2N\) amplitudes and
scalar maps \(N\); a fifth control adds \(\beta_{\rm spectral}=0.1\) to the
vorticity fit.

Three seeded freely decaying realizations use a \(96^2\), \(2\pi\)-periodic
grid, initial energy \(0.5\), and a vorticity spectrum centered at \(k_0=8\).
Each advances 300 RK4 steps with \(\Delta t=5\times10^{-3}\),
\(\nu=2\times10^{-3}\), no forcing or drag, and Orszag \(2/3\) dealiasing.
All maps share centers, widths, and budgets
\(N\in\{32,64,128,256,512\}\). Storage-matched comparisons use
\(D\in\{64,128,256,512\}\), with \(N=D/2\) for vector and \(N=D\) for
scalar amplitudes.

\begin{figure}
  \centering
  \includegraphics[width=\linewidth]
  {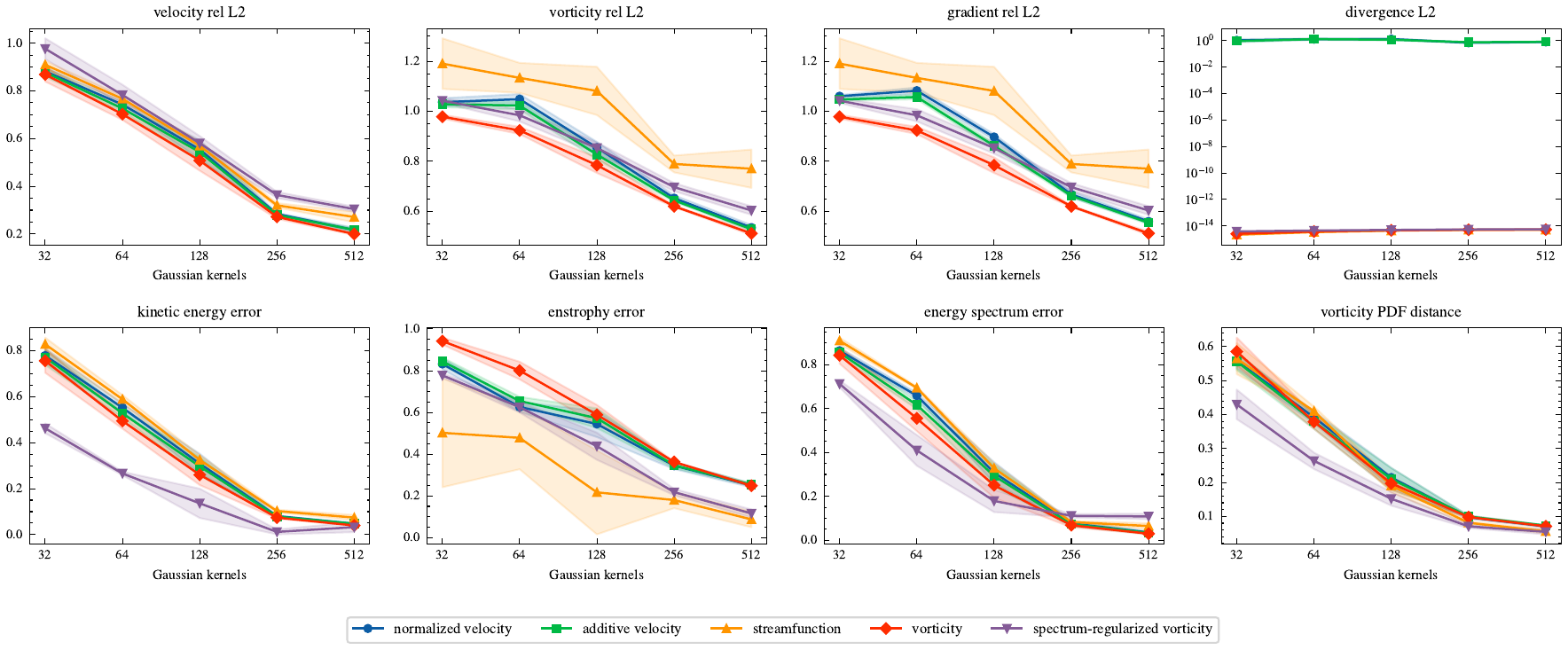}
  \caption{
  Represented-variable control across Gaussian budgets. Curves and bands show
  means and sample standard deviations over three realizations. Vorticity gives
  the strongest storage-matched result; both potential controls preserve
  incompressibility to numerical precision. Lower is better.
  }
  \label{fig:gaussian-budget-sweep}
\end{figure}

\begingroup
Figure~\ref{fig:gaussian-budget-sweep} shows the lowest errors for vorticity at
every budget and the best mean rank across eight diagnostics. Both
potential-based controls are divergence-free to \(O(10^{-15})\), but vorticity
fitting weights Fourier velocity error by \(|\mathbf k|^2\). Spectral
regularization improves kinetic-energy accuracy at the expense of velocity
fidelity. The comparison therefore supports unregularized vorticity for the
three-dimensional curl-lifted study.
\endgroup

\clearpage

\clearpage

\section{Evidence scope and reproducibility}
\label{supp:audit-protocol}

Section~\ref{M-app:relabel-protocol} defines the perturbations, admissibility
gates, and diagnostics. This section separates independently reproducible
controls from conclusions supported only by aggregate records.

\subsection{Gaussian relabelling evidence}

The available Gaussian record reports the declared
\(\tau_{\mathrm{PA}}\) and \(\alpha_{\mathrm{alloc}}\) grids, 1404 total
trials, 509 admitted trials, and 37 admitted trials that break an attribution
criterion. It does not contain the trial-level perturbations or the complete
generation rule. In particular, the perturbation-magnitude grid, Cartesian
product and pruning rule, any random seed, ensemble reduction of the
admissibility gate, exact \(1\%\) denominators, covariance update order,
strongest-trial tie rule, and mode-B normalization remain unspecified.

The article therefore does not use these counts for a binary Gaussian verdict.
Its Gaussian conclusion rests on the independently reproducible
principal-overlap, shared--unique cancellation, and exact block-accounting
results.

\subsection{Independent external-control reproduction}

The coherent--incoherent control uses three periodized wavelet families:
Coiflet 2, Daubechies 4, and Symlet 4. Membership is recomputed for every
snapshot. The integrity-basis control generates 2048 coefficient perturbations
with seed \(20260801\).
Their mathematical definitions and interpretive scope are given in
Sec.~\ref{M-app:relabel-protocol}.

An independent reconstruction of the 128 processed selection snapshots
reproduces the scientific outputs of both controls to relative and absolute
tolerances \(2\times10^{-12}\), including the integrity-basis dominant-term
result. The comparison is restricted to scientific quantities and does not
extend to the unavailable trial-level evidence behind the Gaussian
\(1404/509/37\) counts.

\subsection{Data-role verification}

Independent split verification recovers 31 geometry origins, 16 selection
origins, 16 reserved origins, and one T202 origin. It confirms zero origin
overlap across the geometry, selection, and reserved roles, zero snapshot
overlap among those roles, and identical preprocessing. T202 shares the three
canonical time indices 0, 2000, and 4000, has 199 new time indices, and repeats
no origin--time pair. These checks distinguish snapshot-disjoint sampling from
time-index independence without treating the pooled ensemble as selection-free.

The numerical split specification and verification summary are reported above.
Raw JHTDB fields are not redistributed.

\bibliography{aipsamp}